%% file: K-eigenstructure_assignment.tex
\documentclass[9pt,shortpaper,twoside,web]{ieeecolor}
\usepackage{generic}
\usepackage[backend=bibtex,bibstyle=ieee,sorting=nyt,doi=false]{biblatex}
\AtNextBibliography{\footnotesize}

\usepackage{amsmath,amssymb,amsfonts}
\usepackage{algorithmic}
\usepackage{graphicx}
\usepackage{textcomp}
\usepackage{color}
\input{color.lrt}
\usepackage{tikz}
\usetikzlibrary{calc,shapes}
\usetikzlibrary{positioning,arrows}
\usepackage{pgfplots}
\usetikzlibrary{fit,calc}
\pgfplotsset{compat=1.14}
\usepackage{psfrag}
\usepackage{graphicx}
\usepackage{amsmath,amsfonts,amsxtra}
\usepackage{mathrsfs}
\usepackage{todonotes}
\usepackage{mathtools,leftidx}
\usepackage{calligra}
\usepackage{float}
\usepackage{cleveref}
\crefname{equation}{}{}
\crefrangelabelformat{equation}{({#3}{#1}{#4}--{#5}{#2}{#6})}%
\crefrangelabelformat{subequation}{({#3}{#1}{#4}--{#5}\crefstripprefix{#1}{#2}#6)}%

\usepackage{environ}
\usepackage{bm} 
\usepackage{bbm}
\allowdisplaybreaks[1]

\newcommand{\col}[1]{\operatorname{col}(#1)}

\newcommand{\diag}[1]{\operatorname{diag}(#1)}

\renewcommand{\d}{\mathrm{d}}
\renewcommand{\t}{^{\top}}

\newcommand{\e}{\text{e}}

\newcommand{\uphi}{\underline{\varphi}_n[x]} 
\newcommand{\uphit}{\underline{\varphi}_n\t[x]}

\tikzset{
	state/.style={circle,draw,minimum size=6ex},
	arrow/.style={-latex, shorten >=1ex, shorten <=1ex}}

\newtheorem{thm}{Theorem}
\newtheorem{Lemma}{Lemma}
\newtheorem{rem}{Remark}

\definecolor{mycolor1}{rgb}{0.00000,0.44700,0.74100}%
\definecolor{mycolor2}{rgb}{0.85000,0.32500,0.09800}%
\definecolor{mycolor3}{rgb}{0.92900,0.69400,0.12500}%

\def\BibTeX{{\rm B\kern-.05em{\sc i\kern-.025em b}\kern-.08em
    T\kern-.1667em\lower.7ex\hbox{E}\kern-.125emX}}
\begin{document}
\title{Data-Driven Control of Linear Parabolic Systems using\\ Koopman Eigenstructure Assignment}
\author{Joachim~Deutscher,~\IEEEmembership{Member,~IEEE}
\thanks{J. Deutscher is with the Institute of Measurement, Control and Microtechnology, Ulm University, Germany. (e-mail: joachim.deutscher@uni-ulm.de).
}
}
\maketitle

\begin{abstract}
This paper considers the data-driven stabilization of linear boundary controlled parabolic PDEs by making use of the Koopman operator. For this, a Koopman eigenstructure assignment problem is solved, which amounts to determine a feedback of the Koopman open-loop eigenfunctionals assigning a desired finite set of closed-loop Koopman eigenvalues and eigenfunctionals to the closed-loop system. It is shown that the designed controller only needs a finite number of open-loop Koopman eigenvalues and modes of the state. They are determined by extending the classical Krylov-DMD to parabolic systems. For this, only a finite number of pointlike outputs and their temporal samples as well as temporal samples of the inputs are required resulting in a data-driven solution of the eigenstructure assignment problem. Exponential stability of the closed-loop system in the presence of small Krylov-DMD errors is verified. An unstable diffusion-reaction system demonstrates the new data-driven controller design technique for distributed-parameter systems.
\end{abstract}


\section{Introduction}
\textbf{Background and Motivation.}
Data-driven control is currently a very active research topic with an increasing interest in the control community. This is due to the fact that  using available data directly for the controller design circumvents the time consuming modelling of the dynamical systems. Furthermore, also unmodelled dynamics or even systems, for which a first principle model is hard to obtain, can be dealt with using data-based methods (see \cite{Ho13,Br22} for an overview). Therefore, a lot research directions have been established for data-driven control, where the application of Willems' fundamental lemma (see, e.g., \cite{DeP20}) and the informativity approach (see, e.g., \cite{vaW20}) are currently prominent design techniques. The focus of these methods is mainly on linear finite-dimensional discrete-time systems, but they can also be extended to the nonlinear case. Recently, the \emph{Koopman operator approach} gains more and more attention, because it offers a new paradigm for the data-driven analysis and control (see \cite{Mau20} for an overview). This method is centered around the \emph{Koopman operator}, which describes the dynamics of a system by observables. The latter lift the dynamics of a nonlinear system into an infinite-dimensional space, where a globally linear description with the linear infinite-dimensional Koopman operator is possible. Hence, classical methods from linear infinite-dimensional system theory (see, e.g., \cite{Cu20}) become available for the analysis of nonlinear systems. An inherent feature of this approach is that finite-dimensional approximations of the Koopman operator are attainable directly from data using \emph{dynamic mode decomposition (DMD)} (see \cite{Ku16,Br22}). This results in purely data-driven analysis and synthesis methods. It has been soon recognized that the Koopman operator description of dynamical systems allows to design controllers directly based on data. In particular, the very challenging problem to develop data-driven control methods for nonlinear systems can systematically be dealt with the Koopman operator due to its linear nature. Recent results include MPC in \cite{Kor18a} and optimal control in \cite{Kai21,Gos21,Hua22}. For distributed-parameter systems, which are described by PDEs, methods for data-driven control are mainly based on a finite-dimensional representation of the underlying infinite-dimensional dynamics. With this, established control methods for finite-dimensional systems become available. Thereby, the DMD allows to determine reliable finite-dimensional models for control (see, e.g., \cite{Pe19,Schau21}). 
Recently, a systematic spectral analysis of the Koopman operator for PDEs was presented in \cite{Na20}, which is used to solve identification problems in \cite{Mau21}. However, to the best knowledge of the author, these interesting results have not been utilized for the data-driven control of distributed-parameter systems without recourse to finite-dimensional approximations so far.

\textbf{Contribution.}
In this paper the new spectral Koopman operator framework is applied to the data-driven control of linear parabolic PDEs with boundary control. While an extension to semilinear PDEs seems possible (see the first results in \cite{Na20,Mau21}), the focus is on the linear case, because explicit results can be derived for data-driven control. This already leads to new results and insights, how data can be used for the controller design if the system to be controlled is described by a PDE. The new spectral formulation introduced in \cite{Na20} is shown to be very well taylored for the modal control of parabolic systems, as it is closely related to the well-known modal analysis of PDEs (see, e.g., \cite{Cu20}).  In order to apply modal control to the Koopman operator for parabolic PDEs, a \emph{Koopman eigenstructure assignment problem} is solved in the paper. This amounts to determine a feedback of the open-loop Koopman eigenfunctionals to place a desired finite set of Koopman eigenvalues for the closed-loop system, which is sufficient to stabilize the system. The remaining degrees of freedom after eigenvalue placement can be used to assign the corresponding closed-loop eigenfunctionals. It is verified that the resulting closed-loop eigenfunctionals allow for a series expansion of any bounded observable (i.e., functional). 
For solving the Koopman eigenstructure assignment problem, the underlying infinite-dimensional system dynamics is represented in the coordinates of the Koopman eigenfunctionals. With this, a parametric solution of the posed Koopman eigenstructure assignment problem is presented in the paper. Thereby, the input has not to be taken into account by augmenting the state space (see \cite[Ch. 1]{Mau20} for details), as it is the case for many Koopman-based control methods. It is demonstrated that the resulting controller only requires knowledge of a finite number of Koopman eigenvalues and the Koopman modes of the state. The latter can be obtained from data resulting in a data-driven solution of the posed Koopman eigenstructure assignment problem. To this end, the classical Krylov-DMD  for finite-dimensional systems (see \cite{Row09}) is extended to parabolic systems. This has the advantage that no dictionary (i.e., a set of basis functional) is required for the DMD. As a result, one obtains spatial samples of the dominant Koopman modes for the parabolic system state together with the corresponding Koopman eigenvalues. By considering the large spatial data limit (i.e., an infinite number of spatial samples) it is shown, under what conditions the Koopman modes and the related eigenvalues can be determined exactly. This result is utilized to propose a systematic approach for the data-driven modal analysis of the open-loop system. The proposed Krylov-DMD only needs a finite number of spatial and temporal samples of the spatial-temporal system profile. Since each of these samples is a functional of the state evaluated at a time instant, one directly obtains the required observables to approximate the Koopman operator. They can be collected in experiments, for example by snapshots using an infrared camera or employing a sufficiently large number of heat sensors, as well as from the FEM simulation of models with a complex geometry. This results in a new data-driven approach to directly determine a \emph{finite-dimensional invariant subspace} of a parabolic system containing the dynamics to be changed by feedback control. This is different, from other data-driven approaches, such as \emph{proper orthogonal decomposition (POD)} (see \cite[Ch. 12]{Br22}), where, in general, no invariant subspace is obtained. Furthermore, the POD needs the PDE for the Galerkin projection, which is not necessary for the Koopman approach. The Krylov-DMD leads to errors in the obtained Koopman eigenvalues and modes. Hence, it is verified that the resulting closed-loop system with the obtained data-driven controller remains exponentially stable in the $L_2$-norm provided these errors are sufficiently small. A simple example of an unstable diffusion-reaction system demonstrates the results of the paper and validates the effectiveness of the new data-driven control technique. 

\textbf{Organization.}
The next section introduces the considered problem. For this also some background material on the Koopman operator and its spectral analysis for PDEs from \cite{Na20} is reviewed to keep the paper self-contained and to obtain a sound basis for the subsequently developed data-driven modal control approach. Section \ref{sec:sf} contains the state feedback design and presents a parametric solution of the Koopman eigenstructure assignment problem. In order to obtain a data-driven solution, the Krylov-DMD is extended in Section \ref{sec:ddewp} for the considered class of parabolic systems. The stability of the resulting closed-loop system is investigated in Section \ref{sec:stab} taking the errors originating from Krylov-DMD into account. Section \ref{sec:ex} demonstrates the results of the paper for an unstable diffusion-reaction system. 



\section{Problem formulation}\label{sec:prob}
\textbf{Parabolic System.} Consider the linear \emph{parabolic system}
\begin{subequations}\label{plant}
\begin{align}
 \dot{x}(z,t) &= \rho x''(z,t) + a(z)x(z,t)         &&\label{parasys}\\
      x'(0,t) &= q_0x(0,t) + u_1(t),                  &&\hspace{-1.5cm} t > 0\\ 
      x'(1,t) &= q_1x(1,t) + u_2(t),                  &&\hspace{-1.5cm} t > 0 \label{parasys2}
\end{align}
\end{subequations}
with the state $x(z,t) \in \mathbb{R}$ defined on $(z,t) \in [0,1]\times\mathbb{R}^+_0$ and the inputs $u_i(t) \in \mathbb{R}$, $i = 1,2$. In the paper the system parameters $\rho \in \mathbb{R}^+$, $a \in C[0,1]$, $q_0$ and $q_1$ are \emph{unknown}. However, it is assumed that the \emph{output data}
\begin{equation}
 \mathbbm{D}_{n+1}(x) = \begin{bmatrix}y_0 & y_1 & \ldots & y_n\end{bmatrix} \in \mathbb{R}^{M \times n+1}\label{xdata} 
\end{equation}
with
$y_k = [\langle x(kt_s),c_1\rangle,\ldots,\langle x(kt_s),c_M\rangle]\t$, $k = 0, 1, \ldots,n$, is available for \eqref{plant}. This data is obtained for the open-loop system, i.e., $u_1 = u_2 = 0$. Therein, $\langle x_1, x_2\rangle = \int_0^1x_1(\zeta)\overline{x_2(\zeta)}\d\zeta$ is the \emph{inner product} and $c_i$ is a rectangular weighting function
\begin{equation}\label{cdef}
 c_i(z) = \frac{1}{2\epsilon}\mathbbm{1}_{[z_i-\epsilon,z_i+\epsilon]}(z),\quad \epsilon > 0, z \in [0,1],
\end{equation}
in which $\mathbbm{1}_{[a,b]}$ denotes the \emph{indicator function}, i.e., $\mathbbm{1}_{[a,b]}(z) = 1$, $z \in [a,b]$, and $\mathbbm{1}_{[a,b]}(z) = 0$ elsewhere.
This describes measurements around the points $z_i \in (0,1)$, $i = 1,\ldots,M$, that are subject to temporal sampling with
sampling time $t_s > 0$. 
It is assumed that the length of the spatial domain is known, which, for convenience, is set to $1$, i.e.,  $z \in [0,1]$.

\textbf{Koopman Operator and Generator.}  In order to introduce the Koopman operator, define the state $x(t) = x(\cdot,t)$ and the operator 
\begin{equation}\label{Adef}
 \mathcal{A}h = \rho h'' + ah, \quad h \in D(\mathcal{A}) \subset X
\end{equation}
with $h \in D(\mathcal{A}) = \{h \in H^2(0,1)\,|\, h'(0) = q_0h(0), h'(1) = q_1h(1)\}$ describing the uncontrolled system \crefrange{parasys}{parasys2}, i.e., $u_1 = u_2 = 0$. Then, the corresponding initial boundary value problem (IBVP) can be represented by the abstract IVP
\begin{equation}\label{absivp}
 \dot{x}(t) = \mathcal{A}x(t), \quad x(0) = x_0 \in X, t > 0,
\end{equation}
in the state space $X = L_2(0,1)$ of absolute square Lebesgue integrable functions endowed with the inner product $\langle x_1, x_2\rangle$. Since $-\mathcal{A}$ is a self-adjoint \emph{Sturm-Liouville operator}, the sequence $\{\phi_i,i \in \mathbb{N}\}$ of its eigenvectors form an orthonormal Riesz basis for $X$ w.r.t. real simple eigenvalues (see \cite{Del03}). This implies that $\mathcal{A}$ is a Riesz-spectral operator generating an \emph{analytic} $C_0$-semigroup $\mathcal{T}(t) : \mathbb{R}^+ \mapsto \mathcal{L}(X)$ describing the classical solution of \eqref{absivp}, i.e., $x(t) = \mathcal{T}(t)x_0$, $t \geq 0$. The Koopman operator $\mathscr{K}(t)$ related to \eqref{plant} describes the time evolution of \emph{observables} $g[x]$, $x \in X$. They are assumed to be bounded linear functionals, i.e., $g : X \mapsto \mathbb{C}$, being elements of the infinite-dimensional space $\mathscr{O} = X'$ of observables. 
Since any measurement for \eqref{plant} is a functional, the observables $g[x]$ can be seen as generalized measurements for \eqref{plant}. Obviously, the elements in \eqref{xdata} qualify as observables, because they are bounded linear functionals on $X$.
 Then, the \emph{Koopman operator} for \eqref{absivp} can be introduced by the composition
\begin{equation}\label{Kop}
 \mathscr{K}(t)g[x] = g[\mathcal{T}(t)x], \quad g \in \mathscr{O}, x \in D(\mathcal{A}), t \geq 0,
\end{equation} 
which describes the time evolution of the observables. For each $t \in \mathbb{R}^+$ the map $\mathscr{K}(t): \mathscr{O} \mapsto \mathscr{O}$ is an infinite-dimensional linear operator (see, e.g., \cite{Na20}). The evolution equation for the observables $g[x]$, i.e., $g[x(t)] = \mathscr{K}(t)g[x(0)]$, $t \geq 0$, is
\begin{equation}\label{timeevobsp}
  \d_tg[x(t)] = \mathscr{A}g[x(t)], \quad t > 0, g[x(0)] \in D(\mathscr{A}) \subset \mathscr{O}.
\end{equation}
Therein, $D(\mathscr{A})$ denotes the domain of the \emph{Koopman generator} $\mathscr{A} : D(\mathscr{A}) \mapsto \mathscr{O}$, which is the infinitesimal generator for  $\mathscr{K}(t)$. The latter is defined by
\begin{equation}\label{Kgen}
 \mathscr{A}g[x] = \lim_{t \to 0^+}\frac{\mathscr{K}(t)g[x] - g[x]}{t},
\end{equation}
in which the limit has to be taken in the topology of $\mathscr{O}$.  Then, the domain $D(\mathscr{A})$ is the set of elements $g[x] \in \mathscr{O}$, for which the limit in \eqref{Kgen} exists. The Koopman generator $\mathscr{A}$ can be determined by using the \emph{G\^ateaux differential} $\delta g[x;\mathcal{A}x]$ of the observable $g[x]$ in the direction of the function $\mathcal{A}x$ with
\begin{equation}\label{fablcalc}
 \mathscr{A}g[x] = \delta g[x;\mathcal{A}x] \!=\! \d_{\epsilon}g[x + \epsilon\mathcal{A}x]|_{\epsilon = 0} \!=\! \int_0^1(\delta_xg[x])(\zeta)\mathcal{A}x(\zeta)\d\zeta,
\end{equation}
which has the representation in form of a linear integral operator resulting in a functional. Therein, $\delta_xg[x] = \delta g[x]/\delta x$ denotes the \emph{functional derivative} of $g[x]$ w.r.t. the function $x$, which yields a function. Hence, \eqref{fablcalc} can be seen as a generalization of the Lie-derivative known from finite-dimensions. 
For further details concerning the definition and existence of the Koopman operator and generator for infinite-dimensional systems, the interested reader is referred to \cite{Na20,Mau21}.

\textbf{Eigenvalue Problem for the Koopman Generator.} 
The eigenvalue problem for the Koopman generator $\mathscr{A}$ amounts to determine the \emph{eigenfunctionals} $\varphi_i[x]$, $i \in \mathbb{N}$, w.r.t. the \emph{Koopman eigenvalues} $\lambda_i$ (\emph{K-eigenvalues} for short) as the nontrivial solutions of the \emph{K-eigenvalue problem}
\begin{equation}\label{kewpp}
 \mathscr{A}\varphi_i[x] = \int_0^1(\delta_x\varphi_i[x])(\zeta)\mathcal{A}x(\zeta)\d\zeta = \lambda_i\varphi_i[x],
\end{equation}
$\varphi_i[x] \in D(\mathscr{A})$, $i \in \mathbb{N}$. In order to solve this eigenvalue problem for the considered linear systems, consider an observable $\varphi[x]$ in $\mathscr{O}$, which by the \emph{Riesz representation theorem}  can always be represented in the form
\begin{equation}\label{efktl}
 \varphi[x] = \langle x,\phi\rangle = \int_0^1\! x(\zeta)\overline{\phi(\zeta)}\d\zeta,\quad  \phi \in X,
\end{equation}
(see \cite[Th. A.3.55]{Cu20}). 
From \eqref{fablcalc} with $g[x] = \varphi[x]$ one obtains
\begin{equation}\label{acal1}
 \mathscr{A}\varphi[x] \!= \!\int_0^1\!\!\d_{\epsilon}(x(\zeta) \!+\! \epsilon\mathcal{A}x(\zeta))_{\epsilon = 0}\overline{\phi(\zeta)}\d\zeta \!=\! \int_0^1\!\!\overline{\phi(\zeta)}\mathcal{A}x(\zeta)\d\zeta
\end{equation}
resulting in a linear integral operator so that $\delta_x\varphi[x] = \overline{\phi}$. Letting $\phi$ be the eigenvectors $\phi_i \in D(\mathcal{A}) \subset X$ of $\mathcal{A}$ satisfying $\mathcal{A}\phi_i = \lambda_i\phi_i$, $i \in \mathbb{N}$, i.e., considering $\varphi_i[x] = \langle x,\phi_i\rangle$, and making use of two integrations by parts, it follows  
\begin{equation}\label{ewpfinc}
 \mathscr{A}\varphi_i[x] = \int_0^1\! x(\zeta)\underbrace{\overline{\mathcal{A}\phi_i(\zeta)}}_{= \lambda_i\overline{\phi_i(\zeta)}}\d\zeta
                         = \lambda_i\underbrace{\int_0^1\! x(\zeta)\overline{\phi_i(\zeta)}\d\zeta}_{\varphi_i[x]}
\end{equation}
in view of $\mathcal{A}^* = \mathcal{A}$ as well as $\lambda_i \in \mathbb{R}$. This shows that $\varphi_i[x] = \langle x,\phi_i\rangle$ is an eigenfunctional of $\mathscr{A}$ w.r.t. the eigenvalue $\lambda_i$ that coincide with eigenvalues of $\mathcal{A}$. Since $-\mathcal{A}$ is a self-adjoint Sturm-Liouville operator with a pure discrete point spectrum,  
this completely determines the structure of the spectrum for the corresponding Koopman generator. 
The eigenfunctionals $\varphi_i[x]$ are also the solution of the \emph{K-eigenvalue problem} for the Koopman operator $\mathscr{K}(t)$, because
\begin{equation}\label{Kopeig}
\mathscr{K}(t)\varphi_i[x] = \langle \mathcal{T}(t)x,\phi_i\rangle = \langle x,\underbrace{\mathcal{T}^*(t)\phi_i}_{=\e^{\lambda_it}\phi_i}\rangle =  \e^{\lambda_i t}\varphi_i[x]
\end{equation}
for $\varphi_i[x] \in D(\mathscr{A}), i \in \mathbb{N}$, due to \eqref{Kop} and the \emph{spectral mapping theorem} (see \cite[Col. IV.3.12][]{Eng00,Mau21}). Therein, $\lambda_i$ are called eigenvalues of the Koopman operator $\mathscr{K}(t)$ for convenience.

\textbf{Koopman Eigenstructure Assignment Problem.} On the basis of the last paragraph, it is possible to formulate the \emph{Koopman eigenstructure assignment problem}: Consider the \emph{state feedback controller}
\begin{equation}\label{sf}
 u_i(t) = -\sum_{j=1}^nk^i_j\varphi_j[x(t)], \quad k^i_j \in \mathbb{R}, i = 1, 2,
\end{equation}
in which $\varphi_i[x]$ are the eigenfunctionals \eqref{efktl}. This leads to the closed-loop dynamics
\begin{subequations}\label{kloop}
\begin{align}
 \d_t g[x(t)] &= \int_0^1(\delta_xg[x(t)])(\zeta)(\rho x''(\zeta,t) + a(\zeta)x(\zeta,t))\d\zeta\\
      x'(0,t) &= q_0x(0,t)-\sum_{j=1}^nk^1_j\varphi_j[x(t)]\\
      x'(1,t) &= q_1x(1,t)-\sum_{j=1}^nk^2_j\varphi_j[x(t)]
\end{align}	
\end{subequations}
for any observable $g[x] \in \mathscr{O}$, which, in general, has the form $g[x] = \langle x, \gamma \rangle$, $\gamma \in X$ (see \eqref{plant}, \eqref{timeevobsp}--\eqref{fablcalc}). Find the \emph{state feedback gains} $k^i_j$, $i = 1,2$, $j = 1,\ldots,n$, in \eqref{sf}, that assign a desired self-conjugate and finite set $\sigma_{\text{c}} = \{\tilde{\lambda}_i, i=1,\ldots,n\}$ as K-eigenvalues to the resulting closed-loop system \eqref{kloop}. Thereby, the remaining closed-loop K-eigenvalues should be left unchanged, i.e., $\tilde{\lambda}_i = \lambda_i$, $i > n$. This means that the state feedback \eqref{sf} only modifies the dynamics in the \emph{invariant subspace} spanned by the eigenfunctionals $\varphi_i[x]$, $i = 1,\ldots,n$. The degrees of freedom contained in the feedback gains after eigenvalue assignment have to be used to shape the corresponding closed-loop eigenfunctionals $\tilde{\varphi}_i[x]$, $i = 1,\ldots,n$.

\begin{rem}\label{rem:fineig}
Since it is well-known that $\mathcal{A}$ can only have a finite number of eigenvalues in the closed right half-plane (see, e.g., \cite{Del03}), the considered finite eigenvalue assignment 
is sufficient for the stabilization of \eqref{plant}.	\hfill $\triangleleft$	 
\end{rem}

\section{State Feedback Design}\label{sec:sf}
For solving the Koopman eigenstructure assignment problem posed in the last section, apply the state feedback \eqref{sf} to \eqref{plant}. This leads to the \emph{closed-loop Koopman generator} 
\begin{equation}
 \tilde{\mathscr{A}}g[x] = \int_0^1(\delta_xg[x])(\zeta)\tilde{\mathcal{A}}x(\zeta)\d\zeta, \quad g[x] \in D(\tilde{\mathscr{A}})
\end{equation}
with $\tilde{\mathcal{A}}h = \rho h'' + ah$ and $D(\tilde{\mathcal{A}}) = \{h \in H^2(0,1)\,|\, h'(0) = q_0h(0)-\sum_{i=1}^nk^1_i\langle h,\phi_i\rangle,
h'(1) = q_1h(1)-\sum_{i=1}^nk^2_i\langle h,\phi_i\rangle\} \subset X$. Hence, the corresponding \emph{K-eigenvalue problem} reads
\begin{equation}\label{clewp}
 \tilde{\mathscr{A}}\tilde{\varphi}_i[x] = \int_0^1(\delta_x\tilde{\varphi}_i[x])(\zeta)\tilde{\mathcal{A}}x(\zeta)\d\zeta = \tilde{\lambda}_i\tilde{\varphi}_i[x],
\end{equation}
$\tilde{\varphi}_i[x] \in D(\tilde{\mathscr{A}})$, $i \in \mathbb{N}$, for $\tilde{\mathscr{A}}$ (cf. \eqref{kewpp}). Hence, according to \eqref{efktl} the \emph{closed-loop eigenfunctionals} $\tilde{\varphi}_i[x]$, $i \in \mathbb{N}$, can be considered in the form
\begin{equation}\label{clefktl}
 \tilde{\varphi}_i[x] = \langle x,\tilde{\psi}_i\rangle, \quad i \in \mathbb{N},
\end{equation}
in which $\tilde{\psi}_i \in D(\tilde{\mathcal{A}}) \subset X$. Note that, $\tilde{\psi}_i$, $i \in \mathbb{N}$, are the eigenvectors of the adjoint operator $\tilde{\mathcal{A}}^*$ w.r.t. the eigenvalues $\tilde{\lambda}_i$. Then, \eqref{clewp} takes the explicit form
\begin{equation}\label{clewp2}
 \tilde{\mathscr{A}}\tilde{\varphi}_i[x] = \int_0^1\tilde{\psi}_i(\zeta)\tilde{\mathcal{A}}x(\zeta)\d\zeta = \tilde{\lambda}_i\tilde{\varphi}_i[x],
\end{equation}
$\tilde{\varphi}_i[x] \in D(\tilde{\mathscr{A}})$, $i \in \mathbb{N}$. In order to determine a modal representation of the closed-loop system \eqref{kloop}, consider the expansion
\begin{equation}\label{gexp}
 g[x] = \langle x,\gamma\rangle = \sum_{i = 1}^\infty g_i\varphi_i[x], \quad \gamma \in X,
\end{equation}	
with $g_i = \overline{\langle \gamma,\phi_i \rangle}$ of an observable $g[x] \in \mathscr{O}$, which follows from inserting the expansion $\gamma = \sum_{i=1}^\infty\langle \gamma,\phi_i\rangle\phi_i$ in terms of the eigenvectors of $\mathcal{A}$, i.e., of the open-loop system. Thereby, the interchanging of the inner product with the sum is allowed due to the boundedness of the inner product w.r.t. the second argument. Inserting \eqref{gexp} in \eqref{kloop}, using $\delta_xg[x] = \sum_{i=1}^\infty g_i\phi_i$, 
applying an integration by parts twice and using the fact that $\phi_i$ are the eigenvectors of $\mathcal{A}$ yield
\begin{multline}\label{eigfunc}
 \d_t{\varphi}_i[x(t)] = \lambda_i\varphi_i[x(t)] -b\t_i\sum_{j=1}^nk_j\varphi_j[x(t)], \quad i \in \mathbb{N},
\end{multline}
with $b_i = \rho\col{-\phi_i(0),\phi_i(1)} \in \mathbb{R}^2$ and $k_j = \col{k^1_j,k^2_j} \in \mathbb{R}^2$. In the following the eigenvalues $\lambda_i$, $i \in \mathbb{N}$, are assumed to be ordered decreasingly.
\begin{rem}\label{funcexpand}
Note that \eqref{eigfunc} is a complete representation of the closed-loop dynamics by the eigenfunctionals $\varphi_i[x]$, $i \in \mathbb{N}$, as observables. This follows from 
\begin{equation}\label{xexpa}
	x(t) = \sum_{i=1}^{\infty}\langle x(t),\phi_i\rangle\phi_i = \sum_{i=1}^{\infty}\phi_i\varphi_i[x(t)],
\end{equation}
which is directly implied by the Riesz basis property of the eigenvectors $\mathcal{A}$ (see Section \ref{sec:prob} and \cite[Lem. 3.2.4]{Cu20}). 
\hfill $\triangleleft$	 
\end{rem}
For assigning $n$ closed-loop K-eigenvalues, consider the first $n$ equations of \eqref{eigfunc}. They can be compactly written as
\begin{subequations}\label{varphisys}
\begin{equation}\label{nsys}
 \d_t\underline{\varphi}_n[x(t)] = (\Lambda_n - B_nK)\underline{\varphi}_n[x(t)],
\end{equation}
in which $\underline{\varphi}_n[x] = \col{\varphi_1[x],\ldots,\varphi_n[x]}$, $\Lambda_n = \diag{\lambda_1,\ldots,\lambda_n}$, $B_n = \col{b\t_1,\ldots,b\t_n}$ and $K = [k_1 \;\; \ldots \;\; k_n]$. This is a classical eigenvalue assignment problem for $\Lambda_n - B_nK$, which is solvable if $(\Lambda_n,B_n)$ is controllable (see, e.g., \cite[Ch. 7.1]{KL80}). Since the eigenvalues $\lambda_i$, $i \in \mathbb{N}$, are distinct and $\Lambda_n$ is diagonal, this is satisfied if $B_n$ has no zero row (see \cite{Gi63}). It is well known, that the spectrum of the Sturm-Liouville operator $-\mathcal{A}$ in \eqref{Adef} is bounded from below (see \cite{Del03}). Hence, there only exists a finite number of eigenvalues $\lambda_i$ in the closed right half-plane. Consequently, by choosing $n$ sufficiently large, one can always ensure $\operatorname{Re}\lambda_i < 0$, $i > n$. The remaining dynamics are
\begin{equation}\label{infsys}
 \d_t\underline{\varphi}_{\infty}[x(t)] = \Lambda_{\infty}\underline{\varphi}_{\infty}[x(t)] - B_{\infty}K\underline{\varphi}_n[x(t)]
\end{equation}
\end{subequations}
with $\underline{\varphi}_{\infty}[x]  = \col{\varphi_{n+1}[x],\varphi_{n+2}[x],\ldots}$, $\Lambda_{\infty} = \diag{\lambda_{n+1},\linebreak\lambda_{n+2},\ldots}$, $\lambda_{n+1} > \lambda_{n+2} > \ldots$,  and $B_{\infty} = \col{b\t_{n+1},b\t_{n+2},\ldots}$. In view of \eqref{varphisys} it follows 
\begin{equation}
 \d_t\begin{bmatrix}
  \underline{\varphi}_n[x(t)]\\
  \underline{\varphi}_{\infty}[x(t]
 \end{bmatrix}
 =
 \begin{bmatrix}
 \Lambda_n-B_nK & 0\\
 -B_{\infty}K   & \Lambda_{\infty} 
 \end{bmatrix}
 \begin{bmatrix}
 \underline{\varphi}_n[x(t)]\\
 \underline{\varphi}_{\infty}[x(t]
 \end{bmatrix}
\end{equation}
so that \eqref{sf} only shifts the first $n$ K-eigenvalues of the closed-loop system. Thereby, the remaining open-loop K-eigenvalues are left unchanged. The next theorem shows that this eigenvalue assignment ensures exponential stability of the resulting closed-loop system.
\begin{thm}[Stabilizing state feedback controller]\label{lem:stabl}\hfill
Assume that $\lambda_{\text{max}} = \max_{\lambda \in \sigma(\Lambda_n-B_nK)}\operatorname{Re}\lambda < 0$,
let $\epsilon > 0$ be such that $\alpha = \lambda_{\text{max}} + \epsilon$ satisfies $\lambda_i < \alpha < 0$, $i > n$. 
Then, the closed-loop system resulting from appyling \eqref{sf} to \eqref{plant} is exponentially stable in the $L_2$-norm $\|\cdot\|$, i.e.,
\begin{equation}\label{lem:secoexp}
 \|x(t)\| \leq M_x\e^{\alpha t}\|x(0)\|,\quad t \geq 0, \forall x(0) \in X
\end{equation}
for $M_x \geq 1$.
\end{thm}
\begin{proof}
Introduce the space $\mathcal{X} = \ell_2 = \{(h_1,h_2,\ldots)\,|\, h_i \in \mathbb{C}, \sum_{i=1}^{\infty}|h_i|^2 < \infty\}$ equipped with the inner product $\langle x,y \rangle_{\ell_2} = \sum_{i=1}^{\infty}x_i\overline{y_i}$, which is the Hilbert space of absolutely square summable sequences $(h_i)_{i\in\mathbb{N}}$ composed of scalars $h_i \in \mathbb{C}$. With $\tilde{A}_n = \Lambda_n - B_nK$ the solution of \eqref{nsys} is $\underline{\varphi}_n[x(t)] = \e^{\tilde{A}_nt}\underline{\varphi}_n[x(0)]$, $t \geq 0$, and thus
\begin{equation}\label{fristnest}
 \|\underline{\varphi}_{n}[x(t)]\|^2_{\mathbb{C}^n} \leq c^2_0\e^{2\alpha t}\|\underline{\varphi}_{n}[x(0)]\|^2_{\mathbb{C}^n}, \quad t \geq 0,
\end{equation}
in light of $\|\e^{\tilde{A}_n t}\| \leq c_0\e^{\alpha t}$, $c_0 \geq 1$, $t \geq 0$. Consider the estimates $\e^{\lambda_{n+1} t} > \e^{\lambda_{i} t}$, $i > n+1$, and $|b_i\t K\e^{\tilde{A}_n t}\underline{\varphi}_n[x(0)]| \leq c_1\e^{\alpha t}\|\underline{\varphi}_n[x(0)]\|_{\mathbb{C}^n}$ for $t \geq 0$, $i > n$, and some $c_1 > 0$. Then, it is readily shown that the result
\begin{multline}\label{pfirstest}
 \|\underline{\varphi}_{\infty}[x(t)]\|^2_{\ell_2} \leq 2\e^{2\lambda_{n+1}t}\|\underline{\varphi}_{\infty}[x(0)]\|^2_{\ell_2}\\
 + 2c_1^2\sum_{i=n+1}^{\infty}\int_0^t\e^{2(\lambda_i t + (\alpha -\lambda_i)\tau)}\d\tau\|\underline{\varphi}_n[x(0)]\|^2_{\mathbb{C}^n}
\end{multline}
holds for the solution of \eqref{infsys} by making use of the \emph{parallelogram law} (see, e.g., \cite[Def. A.2.26]{Cu20}). Evaluating the integral gives
\begin{equation}\label{psumconv}
 \sum_{i=n+1}^{\infty}\int_0^t\e^{2(\lambda_i t + (\alpha -\lambda_i)\tau)}\d\tau \leq \sum_{i=n+1}^{\infty}\frac{c_2\e^{2\alpha t}}{2|\alpha - \lambda_i|} = c_3\e^{2\alpha t}
\end{equation}
for some $c_2, c_3 > 0$. This follows from a simple calculation and the fact that $|\lambda_i| \in O(i^2)$ holds for the Sturm-Liouville operator $-\mathcal{A}$ in \eqref{Adef} (see, e.g., \cite{Orl17a}) implying the convergence of the series in \eqref{psumconv}. By $\lambda_i < \alpha$, $i > n$, the estimate $\e^{\alpha t} > \e^{\lambda_{n+1} t}$, $t \geq 0$, holds so that \eqref{pfirstest} and \eqref{psumconv} lead to
$\|\underline{\varphi}_{\infty}[x(t)]\|^2_{\ell_2} \leq c_4\e^{2\alpha t}\|\underline{\varphi}[x(0)]\|^2_{\ell_2}$, $t \geq 0$, for some $c_4 > 0$ in view of $\|\underline{\varphi}[x(t)]\|^2_{\ell_2} = \|\underline{\varphi}_{n}[x(t)]\|^2_{\mathbb{C}^n} + \|\underline{\varphi}_{\infty}[x(t)]\|^2_{\ell_2}$, $t \geq 0$. Combining this with \eqref{fristnest} 
yields 
\begin{equation}\label{varexp}
 \|\underline{\varphi}[x(t)]\|_{\ell_2}\linebreak \leq M\e^{\alpha t}\|\underline{\varphi}[x(0)]\|_{\ell_2}, \quad t \geq 0,  \underline{\varphi}[x(0)] \in \mathcal{X}, 
\end{equation}
$\forall M \geq 1$, after a straightforward computation.
Since the sequence $\{\phi_i, i \in \mathbb{N}\}$ is an orthonormal Riesz basis for $X$ (see \cite{Del03}), the \emph{Parseval's equality} $\|\sum_{i=1}^{\infty}h_i\phi_i\|^2 = \sum_{i=1}^{\infty}|h_i|^2 = \|h\|^2_{\ell_2}$ is satisfied. Then, \eqref{xexpa} and \eqref{varexp} directly imply \eqref{lem:secoexp}. 
\end{proof}

The usual approach to solve the eigenstructure assignment problem for boundary controlled systems (see, e.g., \cite{Ko75,Deu09}), which directly uses the closed-loop eigenvalue problem for determining the state feedback gains, does not work for Koopman eigenstructure assignment. 
This is due to the fact that in \cite{Ko75,Deu09} the eigenvectors of the closed-loop system operator are utilized to determine the state feedback gains. In contrast, the eigenfunctional \eqref{clefktl} contains the eigenvector $\tilde{\psi}_i$, which is related to the adjoint operator $\tilde{\mathcal{A}}^*$ (see \eqref{clefktl}). This leads to new challenges for the Koopman eigenstructure assignment problem, which is solved in the next theorem.
\begin{thm}[Parametric state feedback gain]\label{theo:para}
Consider the system (\ref{varphisys}) and a given self-conjugate set of different numbers $\{\tilde{\lambda}_i, i=1,\ldots,n\} \notin \sigma(\Lambda_n) \cup \sigma(\Lambda_{\infty})$, $\tilde{\lambda}_i \in \mathbb{C}$, with corresponding \emph{parameter vectors} $p_i \in \mathbb{C}^2$, $i = 1,\ldots,n$. Further assume that the $\tilde{\lambda}_{i}$ and the $p_{i}$ have been chosen such that the vectors 
\begin{equation}
 \tilde{v}_i = (\Lambda_n - \tilde{\lambda}_iI)^{-1}B_np_i, \quad i = 1,\ldots,n,
\end{equation}
are linearly independent. Then, the feedback gain 
\begin{equation}\label{Kfinal}
	K = \begin{bmatrix}p_1 & \ldots & p_n\end{bmatrix}
	\begin{bmatrix}\tilde{v}_1 & \ldots & \tilde{v}_n\end{bmatrix}^{-1} = P\tilde{V}^{-1}
\end{equation}
assigns the K-eigenvalues $\tilde{\lambda}_i$, $i = 1,\ldots,n$, and the parameter vectors $p_i$, $i = 1,\ldots,n$, defined by $p_i = K\tilde{v}_i$ to the closed-loop system. The $n$ degrees of freedom remaining in $K$ after eigenvalue assignment can be used to influence the closed-loop eigenfunctionals according to 
\begin{align}\label{tilvarphipar}
  \tilde{\varphi}_i[x] = 
  \begin{cases}
   \underline{c}_i\t\underline{\varphi}_n[x], & i = 1,\ldots,n\\
   \underline{c}_i\t\underline{\varphi}_n[x] + \varphi_i[x], & i > n
  \end{cases},
\end{align}
in which $\underline{c}_i \in \mathbb{C}^n$, $\underline{\varphi}_n[x] = \col{\varphi_1[x],\ldots,\varphi_n[x]}$ (cf. \eqref{efktl}),
\begin{align}\label{cleigfktl}
 \bar{\underline{c}}_i\t = 
 \begin{cases}
  e_i\t\tilde{V}^{-1}, & i = 1,\ldots,n\\
  b\t_{i}P(\tilde{\Lambda}_n - \lambda_iI)^{-1}\tilde{V}^{-1}, & i > n,
 \end{cases}
\end{align}
where $e_i$, $i = 1,\ldots,n$, being the unit vectors in $\mathbb{R}^n$ and $\tilde{\Lambda}_n = \diag{\tilde{\lambda}_1,\ldots,\tilde{\lambda}_n}$.
\end{thm}
\begin{proof}
By making use of the eigenvalue problem for \eqref{Adef}, it is easily verified that always $\phi_i(0) \neq 0$ and $\phi_i(1) \neq 0$, $i \in \mathbb{N}$. This implies that $B_n$ in \eqref{nsys} has no zero row. With this, the result for the state feedback gain directly follows from the corresponding result in \cite{Deu09}. In order to determine the closed-loop eigenfunctionals, insert $\tilde{\varphi}_i[x] = \sum_{j=1}^{\infty}c_{ij}\varphi_j[x]$ in \eqref{clewp}. With the same calculations to obtain \eqref{eigfunc}, this gives
\begin{equation}
 \sum_{j=1}^{\infty}\overline{c_{ij}}(\lambda_j\varphi_j[x]
 -b\t_j\sum_{l=1}^nk_l\varphi_l[x]) = \tilde{\lambda}_i\sum_{j=1}^{\infty}\overline{c_{ij}}\varphi_j[x]
\end{equation}
for $i \in \mathbb{N}$. By introducing $\underline{c}_{i,n}  = \col{c_{i1},\ldots,c_{in}}$ and $\underline{c}_{i,\infty}  = \col{c_{i,n+1},c_{i,n+2},\ldots}$ the result
\begin{subequations}
\begin{align}
 \bar{\underline{c}}_{i,n}\t(\Lambda_n - B_nK) - \bar{\underline{c}}_{i,\infty}\t B_{\infty}K &= \tilde{\lambda}_i\bar{\underline{c}}_{i,n}\t\label{infeq}\\
                            \bar{\underline{c}}_{i,\infty}\t\Lambda_{\infty} &= \tilde{\lambda}_i\bar{\underline{c}}_{i,\infty}\t\label{infeig}
\end{align} 
\end{subequations}
directly follows for $i \in \mathbb{N}$ in view of \eqref{varphisys}. With $I_{\infty}$ denoting the identity in $\ell_2$, the equation \eqref{infeig} can be rewritten as $\bar{\underline{c}}_{i,\infty}\t(\Lambda_{\infty} - \tilde{\lambda}_iI_{\infty}) = 0\t$ so that $\bar{\underline{c}}_{i,\infty}\t = 0\t$ due to  $\tilde{\lambda}_i \notin \sigma(\Lambda_\infty)$, $i = 1,\ldots,n$. Consequently, $\bar{\underline{c}}_{i,n}\t$ are the left eigenvectors of $\Lambda_n - B_nK$ w.r.t. $\tilde{\lambda}_i$, i.e., $\bar{\underline{c}}_{i,n}\t(\Lambda_n - B_nK) = \tilde{\lambda}_i\bar{\underline{c}}_{i,n}\t$, $i = 1,\ldots,n$, in view of \eqref{infeq}. In \cite{Deu09} it is verified that $K$ in \eqref{Kfinal} assigns the modal matrix $\tilde{V}$ to $\Lambda_n - B_nK$. Since the rows of $\tilde{V}^{-1}$ coincide with $\bar{\underline{c}}_{i,n}\t$, the result \eqref{cleigfktl} for $i = 1,\ldots,n$ is shown. In the case $i > n$ with $\tilde{\lambda}_i = \lambda_i$ the equation \eqref{infeig} yields $c_{ii} \neq 0$ arbitrary and $c_{ij} = 0$, $j \neq i$, which follows from a direct calculation. Hence, by choosing $c_{ii} = 1$ the result $\bar{\underline{c}}\t_{i,n} = b_i\t K(\Lambda_n - B_nK - \lambda_iI)^{-1}$ is obtained from \eqref{infeq}, where the inverse exists by the assumption $\tilde{\lambda}_i \notin \sigma(\Lambda_n)$. Using $\tilde{V}^{-1}(\Lambda_n - B_nK)\tilde{V} = \tilde{\Lambda}_n$  the result \eqref{cleigfktl} for $i > n$ can readily be verified.
\end{proof}	
An important issue in the eigenstructure assignment is the property that the closed-loop eigenfunctionals $\tilde{\varphi}_i[x] = \langle x,\tilde{\psi}_i\rangle$, $i \in \mathbb{N}$, can be used to represent any analytic observable $g[x] \in \mathcal{O}$. This, in particular, requires that the sequence
\begin{align}\label{tilphipar}
\tilde{\psi}_i(z) = 
\begin{cases}
\underline{c}_i\t\underline{\phi}_n(z), & i = 1,\ldots,n\\
\underline{c}_i\t\underline{\phi}_n(z) + \phi_i(z), & i > n,
\end{cases}
\end{align}
and $\underline{\phi}_n = \col{\phi_1,\ldots,\phi_n}$ directly resulting from \eqref{clefktl} and \eqref{tilvarphipar} is a Riesz basis for $X$. The next lemma presents the corresponding conditions.
\begin{Lemma}[Riesz basis assignment]\label{rbasistheo}
Assume that the closed-loop eigenvalues $\tilde{\lambda}_i$, $i = 1,\ldots,n$, are distinct and are not contained in $\sigma(\Lambda_{\infty})$. Then, the sequence $\{\tilde{\psi}_{i}, i \in \mathbb{N}\}$ in \eqref{tilphipar} is a Riesz basis for $X$.
\end{Lemma}
\begin{proof}	
The sequence of vectors $\{\tilde{\psi}_{i}, i \in \mathbb{N}\}$ are linearly independent and thus also \emph{$\omega$-linearly independent} in $X$ (see \cite[Ch. 1.1]{Guo19}). This follows from the fact that all closed-loop eigenvalues are distinct (see \cite[Th. 7.4-3]{Kr78}), because the spectrum $\sigma(\Lambda_{\infty})$ is discrete with simple eigenvalues (see Section \ref{sec:prob}), $\{\tilde{\lambda}_i, i=1,\ldots,n\} \notin \sigma(\Lambda_{\infty})$ and distinct closed-loop eigenvalues $\tilde{\lambda}_i$, $i = 1,\ldots,n$. Next, it is shown that $\{\tilde{\psi}_{i}, i \in \mathbb{N}\}$ is \emph{quadratically close} to the orthonormal basis $\{\phi_i, i \in \mathbb{N}\}$ (see Section \ref{sec:prob}), i.e.,
$\sum_{i=1}^{\infty}\|\tilde{\psi}_i -\phi_i\|^2 < \infty$ holds. For this consider $\sum_{i=1}^{\infty}\|\tilde{\psi}_i -\phi_i\|^2 = \sum_{i=1}^{n}\|e_i\t\tilde{V}^{-1}\underline{\phi}_n - \phi_i\|^2 
 + \sum_{i=n+1}^{\infty}\|b_i\t P(\tilde{\Lambda}_n - \lambda_iI)^{-1}\tilde{V}^{-1}\underline{\phi}_n\|^2$ in view of \eqref{cleigfktl} and \eqref{tilphipar}. Therein,  $\tilde{V}^{-1}$ exists, because the closed-loop eigenvalues $\tilde{\lambda}_i$, $i=1,\ldots,n$, are distinct (see \cite{Deu09}). This sum can be estimated by $\sum_{i=1}^{\infty}\|\tilde{\psi}_i -\phi_i\|^2 \leq \kappa_1 + \sum_{i=n+1}^{\infty}\sum_{j=1}^n\|\frac{b_i\t p_j\tilde{w}_i\t}{\tilde{\lambda}_j - \lambda_i}\|^2 
\leq \kappa_1 + \kappa_2\sum_{i=n+1}^{\infty}\sum_{j=1}^n\frac{1}{|\tilde{\lambda}_j - \lambda_i|^2}$ for some positive constants $\kappa_i$, $i = 1,2$, and $\tilde{V}^{-1} = \col{\tilde{w}\t_1,\ldots,\tilde{w}\t_n}$. The sum at the right hand side is convergent, because $|\lambda_i| \in O(i^2)$ holds for the Sturm-Liouville operator $-\mathcal{A}$ in \eqref{Adef} (see, e.g., \cite{Orl17a}). Hence, the sequence $\{\tilde{\psi}_i, i\in \mathbb{N}\}$ is quadratically close to $\{\phi_i, i\in \mathbb{N}\}$ so that by \emph{Bari's theorem} the $\omega$-linearly independent sequence $\{\tilde{\psi}_{i}, i \in \mathbb{N}\}$ is a Riesz basis for $X$ (see, e.g., \cite[Th. 2.7]{Guo19}).
\end{proof}
Note that the condition of Lemma \ref{rbasistheo} can always be satisfied in the design if $(\Lambda_n,B_n)$ is controllable. Based on the result of Lemma \ref{rbasistheo}, it is possible to obtain a representation for $x$ in terms of the closed-loop eigenfunctionals $\tilde{\varphi}_i[x]$, $i \in \mathbb{N}$. Observe that $x(z,t)$, i.e., the evaluation of the state at $(z,t)$, is a functional and thus an observable. Hence, it is possible to expand the state by $\tilde{\varphi}_i[x]$ but with spatially dependent coefficients.  For this, observe that there always exists a unique \emph{biorthonormal sequence} $\{\tilde{\phi}_i, i \in \mathbb{N}\}$, i.e., $\langle \tilde{\psi}_i,\tilde{\phi}_j\rangle = \delta_{ij}$, $i,j \in \mathbb{N}$, such that
\begin{equation}\label{xexpa2}
 x(t) = \sum_{i=1}^{\infty}\langle x(t),\tilde{\psi}_i\rangle\tilde{\phi}_i = \sum_{i=1}^{\infty}\tilde{\phi}_i\tilde{\varphi}_i[x(t)]
\end{equation}
(see \cite[Lem. 3.2.4]{Cu20}). Then, any bounded linear functional in $\mathscr{O}$ has the expansion
\begin{equation}
 g[x] = \langle x,\gamma\rangle = \sum_{i = 1}^\infty \tilde{g}_i\tilde{\varphi}_i[x], \quad \gamma \in X,
\end{equation}	
with the \emph{Koopman modes} $\tilde{g}_i = \overline{\langle \gamma, \tilde{\psi}_i\rangle}$ (cf. Remark \ref{funcexpand}). 

\section{Krylov-DMD for Parabolic Systems}\label{sec:ddewp}
In what follows the Krylov-DMD formulated for ODEs (see, e.\:g. \cite{Row09,Ar17,Dr18,Me22}) is extended to the parabolic system \eqref{plant}. 
This allows to determine the Koopman eigenvalues and  modes of $x$ from the available data so that also the corresponding eigenfunctionals $\varphi_i[x]$ and the input vectors $b_i$ in \eqref{nsys} can be computed. 

Introduce the vector valued observable $d[x] = \langle x,c\rangle$, in which $c(z) = \col{c_1(z),\ldots,c_M(z)} \in \mathbb{R}^M$. 
Consider
\begin{equation}\label{lkf}
 \mathscr{K}^n(t_s)d[x(0)] = - \sum_{k=0}^{n-1}f_k\mathscr{K}^k(t_s)d[x(0)] + r_{n}.
\end{equation}
The error $r_{n}$ will not vanish if $\mathscr{K}^n(t_s)d[x(0)] \notin \operatorname{span}(d[x(0)],\linebreak\mathscr{K}(t_s)d[x(0)],\ldots,\mathscr{K}^{n-1}(t_s)d[x(0)])$. Then, however, it is still possible to determine $f_k$, $k = 0,\ldots,n-1$, such that the norm of $r_{n}$ is minimal. As a result, a finite-dimensional approximation of the Koopman operator $\mathscr{K}$ in form of a finite-dimensional matrix is obtained (see \cite{Me22}). This leads to the \emph{companion-matrix DMD} presented in \cite{Row09,Ar17}. For this, use \eqref{lkf} to obtain
\begin{equation}\label{Kdmd}
  \mathscr{K}(t_s)\mathbbm{D}_{n}(x) =  
  \mathbbm{D}_{n}(x)
  \!\underbrace{\begin{bsmallmatrix}
  	0      &  0      & \ldots & 0   & -f_0\\
  	1      &  0      & \ldots & 0   & -f_1\\
  	0      &  1      & \ldots & 0   & -f_2\\
  	\vdots &  \vdots & \ddots & 0   & \vdots\\
  	0      &  0      & \ldots & 1   & -f_{n-1}
  	\end{bsmallmatrix}}_{F}  + \, r_{n}e_{n}\t
\end{equation}
with \eqref{xdata}  and the \emph{companion matrix} $F \in \mathbb{R}^{n \times n}$. The norm of the residual $r_n \in \mathbb{R}^M$ becomes minimal for
\begin{equation}\label{fdet}
 f = \begin{bmatrix} f_0 & \ldots & f_{n-1}\end{bmatrix}\t = -\mathbbm{D}^+_{n}(x)x_n,
\end{equation}
in which $\mathbbm{D}^+_{n}(x)$ is the \emph{Moore-Penrose inverse} (see, e.g., \cite[Ch. 12.8]{La85}). If $r_n = 0$ in \eqref{Kdmd}, then the span of the columns in $\mathbbm{D}_{n}(x)$ is invariant w.r.t. the action of $\mathscr{K}(t_s)$ so that in this case $F$ is a finite-dimensional representation of the Koopman operator on this subspace. Hence, for $r_n \neq 0$ the matrix $F$ can be seen as a finite-dimensional approximation of the Koopman operator $\mathscr{K}(t_s)$ (see \cite{Me22}). Therefore, approximations of the Koopman eigenvalues and the Koopman modes of $x$ can be determined by solving an eigenvalue problem for $F$. To this end, 
assume that $F$ has only the simple eigenvalues $\mu_i$, $i = 1,\ldots,n$, with associated eigenvectors $v_i \in \mathbb{R}^n$. Then, postmultiplying \eqref{Kdmd} by $v_i$ yields
\begin{align}\label{Kdmdm}
\mathscr{K}(t_s)\mathbbm{D}_{n}(x)v_i &=  
\mathbbm{D}_{n}(x)Fv_i +  r_{n}e_{n}\t v_i\nonumber\\ 
&= \mu_i\mathbbm{D}_{n}(x)v_i +  r_{n}e_{n}\t v_i, 
\end{align}
for $i = 1,\ldots,n$. The next lemma shows that \eqref{Kdmdm} allows to determine the Koopman eigenvalues and the Koopman modes of the state $x$.
\begin{thm}[Companion-matrix DMD]\label{thm:krylm}
Assume that $F$ has only the simple eigenvalues $\mu_i$, $i = 1,\ldots,n$, with associated eigenvectors $v_i \in \mathbb{R}^n$. Then, 
\begin{equation}\label{phiappm}
 \mathbbm{D}_{n}(x)v_i = \begin{bmatrix}\hat{\phi}_i(z_1) & \ldots & \hat{\phi}_i(z_M)\end{bmatrix}\t, \quad i = 1,\ldots,n,
\end{equation}
where $\hat{\phi}_i$ is an approximation of the eigenvector $\phi_i$, $i = 1,\ldots,n$, w.r.t. the eigenvalue $\e^{\lambda_it_s}$ of  $\mathcal{T}(t_s)$ in the sense of the \emph{weighted residual}
\begin{equation}\label{res}
 \langle\underbrace{\mathcal{T}(t_s)\hat{\phi}_i - \mu_i\hat{\phi}_i}_{= \tilde{r}_ne_n\t v_i},c\rangle = r_ne\t_{n}v_i
\end{equation}%
with the $C_0$-semigroup $\mathcal{T}(t)$, $t \geq 0$, of \eqref{plant} and $r_n = \langle \tilde{r}_n,c\rangle$. 
Consequently, \eqref{phiappm} is an approximation  of the Koopman modes for $x$ (cf. \eqref{efktl})  w.r.t. the approximate Koopman eigenvalues $\hat{\lambda}_i = \ln\mu_i/t_s$, where the residual $\tilde{r}_n$ is minimal around the sampling points $z_i$, $i = 1,\ldots,M$ for small $\epsilon$ (cf. \eqref{cdef}.
\end{thm}
\begin{proof}	
In view of the spectral representation $\mathcal{T}(t_s)h  = \sum_{i=1}^{\infty}\langle h,\phi_i\rangle\e^{\lambda_it_s}\phi_i$ (see \cite[Th. 3.2.8]{Cu20}) it directly follows that $\mathcal{T}(t_s)\phi_i = \e^{\lambda_it_s}\phi_i$, $i \in \mathbb{N}$. Hence, the Koopman modes $\phi_i$ of $x$ are the eigenvectors of $\mathcal{T}(t_s)$ w.r.t. the Koopman eigenvalues $\lambda_i$ (cf. \eqref{Kopeig} and \eqref{xexpa}). After defining
$\mathbbm{d}\t_{n}(x) = [x(0), x(1), \ldots, x(n-1)] \in X^{1 \times n}$ it is readily verified that
\begin{equation}\label{eigd}
  \mathbbm{D}_{n}(x)v_i = \langle \mathbbm{d}\t_{n}(x)v_i, c\rangle = \langle \hat{\phi}_i,c\rangle.
\end{equation}
Then, \eqref{Kdmdm} gives \eqref{res}. 
Obviously, $\hat{\phi}_i = \mathbbm{d}\t_{n}(x)v_i$ is an approximation of the eigenvector $\phi_i$ of $\mathcal{T}(t_s)$ w.r.t. $\mu_i$, where $\tilde{r}_n$ is minimized at the spatial sampling points by \eqref{fdet}. Then, $\hat{\lambda}_i = \ln\mu_i/t_s$ are the corresponding eigenvalue approximations. With this and \eqref{eigd} the result \eqref{phiappm} directly follows.
\end{proof}	
Once the Koopman modes have been found, also the corresponding \emph{approximate Koopman eigenfunctionals} 
$\hat{\varphi}_i[x] = \langle x,  \hat{\phi}_i\rangle$, $i = 1,\ldots,n$, are known  (cf. \eqref{efktl}). Therein, $\hat{\phi}_i$ is a function resulting from a suitable interpolation of \eqref{phiappm}. 
\begin{rem}	
It is well-known that the Krylov-DMD is not numerically well-posed due to the application of the Krylov sequences for analyzing the data (see, e.g., \cite[Ch. 7]{Mau21} for a discussion). This, however, is not a serious issue for the considered setup, since it is sufficient  to only determine a few dominant modes for control. Nevertheless, it is possible to directly apply the SVD-enhanced Krylov-DMD methods presented in \cite{Ar17} if a numerical stable algorithm is required. 
\hfill $\triangleleft$	 
\end{rem}

In \cite{Me22} pseudospectral convergence is verified for the Krylov-DMD applied to finite-dimensional systems. Though a similar results seems possible for stable parabolic systems, the required large data limit is not attainable in practice for the considered sequential data. This is due to the fact that the spectrum of $\mathcal{A}$ in \eqref{Adef} growths with  $|\lambda_i| \in O(i^2)$. Consequently, only the first dominant modes will be significantly present in the data and can thus only be captured by the Krylov-DMD given a finite numerical accuracy. Further insight in the Krylov-DMD for parabolic systems is gained by considering the large spatial data limit, i.e., $M \to \infty$. Then, the Koopman modes can directly be obtained from the data $\mathbbm{d}\t_{n}(x)$ (see proof of Theorem \ref{thm:krylm}). The next lemma shows that the exact K-eigenvalues and K-modes are obtained with the Krylov-DMD if the initial state of \eqref{plant} can be described by a finite number of K-eigenfunctionals.
\begin{Lemma}[Exact companion-matrix DMD]\label{exactdmd}
Assume that 
\begin{equation}\label{icfin}
 x(0) = \sum_{i=1}^n\phi_i\varphi_i[x(0)],
\end{equation}
in which $\varphi_i[x(0)] \neq 0$, $i = 1,\ldots,n$ and $\phi_i$ are the eigenvectors of $\mathcal{A}$ w.r.t. the eigenvalues $\lambda_i$ in \eqref{Adef}. Let $v_i$, $i = 1,\ldots,n$, be the eigenvectors of  $F$ w.r.t. the eigenvalues $\mu_i$. Then, 
\begin{equation}\label{phiappe}
 \mathbbm{d}\t_{n}(x)v_i = \varphi_i[x(0)]\phi_i, \quad i = 1,\ldots,n,
\end{equation}
(see proof of Theorem \ref{thm:krylm}) and $\mu_i = \e^{\lambda_i t_s}$. 
\end{Lemma}	
\begin{proof}	
By defining $\Phi_n = [\phi_1,\ldots,\phi_n]$ and $\underline{\varphi}_n[x(0)] = \col{\varphi_1[x(0)],\ldots,\varphi_n[x(0)]}$ the IC \eqref{icfin} can be represented by $x(0) = \Phi_n\underline{\varphi}_n[x(0)]$. Then, in view of $x(k) = \mathcal{T}^k(t_s)x(0), k \in \mathbb{N}$, and $\mathcal{T}^k(t_s)\phi_i = \e^{\lambda_ikt_s}\phi_i$, $k \in \mathbb{N}$, the result $x(k)= \Phi_n\Lambda_n^k\underline{\varphi}_n[x(0)]$ with $\Lambda_n = \diag{\e^{\lambda_1t_s},\ldots,\e^{\lambda_nt_s}}$ readily follows. The Cayley-Hamilton theorem applied to $\Lambda_n$ yields $p(\Lambda_n) = \Lambda_n^n + a_{n-1}\Lambda_n^{n-1} + \ldots + a_1\Lambda_n + a_0 I = 0$ with $p(s) = \det(sI-\Lambda_n)$ so that $x(n) = \mathcal{T}^n(t_s)x(0) = \Phi_n\Lambda_n^n\underline{\varphi}_n[x(0)] = -\sum_{i=0}^{n-1}a_i\Phi_n\Lambda_n^i\underline{\varphi}_n[x(0)] = -\sum_{i=0}^{n-1}a_ix(i)$. Consequently, the result $\mathcal{T}(t_s)\mathbbm{d}\t_{n}(x) = \mathbbm{d}\t_{n}(x)F$ holds, i.e., $\tilde{r}_n = 0$ in $\mathcal{T}(t_s)\hat{\phi}_i = \e^{\hat{\lambda}_it_s}\hat{\phi}_i + \tilde{r}_{n}e\t_{n}v_i$, implying $f_i = a_i$, $i = 0,\ldots,n-1$ and thus $\det(sI - F) = p(s)$. This verifies $\mu_i = \e^{\lambda_i t_s}$, $i = 1,\ldots,n$. In order to show \eqref{phiappe}, consider
\begin{align}
 &\mathbbm{d}\t_{n}(x)v_i = \begin{bmatrix}
  x(0) & x(1) & \ldots & x(n-1)
 \end{bmatrix}v_i\nonumber\\
  &= \Phi_n\begin{bmatrix}
  \underline{\varphi}_n[x(0)] & \Lambda_n\underline{\varphi}_n[x(0)] & \ldots & \Lambda_n^{n-1}\underline{\varphi}_n[x(0)]
 \end{bmatrix}v_i.
\end{align}
In view of $\Lambda_n^k\underline{\varphi}_n[x(0)] = \diag{\varphi_1[x(0)],\ldots,\varphi_n[x(0)]}\col{\e^{\lambda_1kt_s},\linebreak\ldots,\e^{\lambda_nkt_s}}$,
$k \in \mathbb{N}$, the result 
\begin{equation}\label{moddet}
 \mathbbm{d}\t_{n}(x)v_i = \Phi_n \diag{\varphi_1[x(0)],\ldots,\varphi_n[x(0)]}V_Fv_i
\end{equation}
is easily obtained, in which $V_F = V_F(\e^{\lambda_1t_s},\ldots,\e^{\lambda_nt_s})$ is the \emph{Vandermonde matrix}. Since the Vandermonde matrix $V_F$ is the inverse modal matrix of $F$ (see, e.g., \cite[Fact 7.18.9]{Bern18}), the result \eqref{phiappe} is implied by $V_Fv_i = e_i$ and \eqref{moddet}.
\end{proof}		
\begin{rem}\label{rem:nchoose}
It is crucial to remark, that the result of Lemma \ref{exactdmd} directly lends itself to a systematics for determining reliable open-loop eigenpairs $(\hat{\varphi}_i[x],\hat{\lambda}_i)$ using Krylov-DMD. In particular, \eqref{icfin} holds with an arbitrarily small error for any $x(0) \in X$ provided that $n$ is sufficiently large. Hence, one can determine the open-loop K-eigenvalues and their eigenfunctionals by successively increasing $n$ until the equation error $r_n$ becomes sufficiently small. Thereby, the equation error $r_n = \mathbbm{D}_{n}(x)f + x_n$ can be evaluated based on known data. This indicates that \eqref{icfin} is nearly satisfied so that a good result can be expected with Theorem \ref{thm:krylm} in view of Lemma \ref{exactdmd}.  \hfill $\triangleleft$	 
\end{rem}
The \emph{approximate input matrix} $\hat{B}_n = \col{\hat{b}\t_1,\ldots,\hat{b}\t_n}$ in \eqref{nsys} with $\hat{b}_i = \hat{\rho}\col{-\hat{\phi}_i(0),\hat{\phi}_i(1)}$ requires to determine $\hat{\rho}$. This can be readily achieved by extending the presented results. Assume that the \emph{step input} $u_2(t) = u_0 s(t)$, $u_0 \neq 0$, and $u_1 = 0$ is applied to the system \eqref{plant} and $0 \notin \sigma(\mathcal{A})$ (cf. \eqref{Adef}). The latter condition is not restrictive, because it can be verified using the previous approach and if necessary an exponential input can be chosen. This gives rise to the \emph{input generator}
\begin{subequations}\label{ugen}
\begin{align}
   \dot{v}(t) &= 0,     && t > 0, v(0) = u_0 \in \mathbb{R}\label{ugenS}\\
       u_2(t) &= v(t),  && t \geq 0.
\end{align}	
\end{subequations}
With this, the extended data matrix
\begin{equation}
 \tilde{\mathbbm{D}}_{n+2}(x) = \begin{bmatrix}u_0     & u_0     & \ldots & u_0\\
                                               d[x(0)] & d[x(1)] & \ldots & d[x(n+1)]
                                \end{bmatrix} 
\end{equation}
is available for the Krylov-DMD obtained from the composite system \eqref{plant} and \eqref{ugen}. Then, approximations for the eigenvalues $\lambda_1 = 0$, $\lambda_i \in \sigma(\mathcal{A})$, $i > 1$, of the extended system \eqref{plant} and \eqref{ugen} can be obtained. It is readily verified that the corresponding Koopman modes are $\psi_{e,i} = \col{0,\psi_i}$, $i > 1$, in which $\psi_i$, $i > 1$, are the eigenvectors of $\mathcal{A}$. In order to get an approximation for $\rho$ using the determined Koopman modes and eigenvalues, consider $ \langle \rho\psi_1'' + a\psi_1, \psi_j\rangle = 0$ for some $j > 1$, which readily follows from the corresponding eigenvalue problem w.r.t. the eigenvalue $\lambda_1 = 0$ yielding the eigenvector $\psi_{e,1} = \col{\psi_{u,1},\psi_1}$. A twofold integration by parts directly yields $\langle \psi_1,\rho\psi_j'' + a\psi_j\rangle + \rho\psi_{u,1} = \lambda_j\langle\psi_1,\psi_j\rangle + \rho\psi_{u,1} = 0$. Hence, 
\begin{equation}
 \rho = - \lambda_j\frac{\langle \psi_1,\psi_j\rangle}{\psi_{u,1}}, \quad j > 1, 
\end{equation}
can be utilized to determine an approximation for $\rho$. Note that $\psi_{u,1} \neq 0$, since $\psi_{u,1}$ is an eigenvector of \eqref{ugenS}. Furthermore, $\langle \psi_1,\psi_j\rangle \neq 0$, $j > 1$, since the sequence $\{\psi_i, i > 1\}$ is an orthonormal basis for $L_2(0,1)$. 
This approach can easily be extended to also consider more general input generators. 

Consequently, the state feedback controller \eqref{sf}, \eqref{Kfinal} can be solely determined from the data.

\section{Stability of the Data-Driven Closed Loop}\label{sec:stab}
The last section shows that it is possible to determine the Koopman eigenvalues $\hat{\lambda}_i$, $i = 1,\ldots,n$, and their eigenfunctionals $\hat{\varphi}_i[x] = \langle x,\hat{\phi}_i\rangle$ from the data \eqref{xdata} using Krylov-DMD. Then, however, the \emph{Krylov-DMD errors} 
$\epsilon_{\lambda} = \|\Lambda_n - \hat{\Lambda}_n\| = \|\Delta\Lambda_n\|$, $\epsilon_i(z) = |\phi_i(z) - \hat{\phi}_i(z)| = |\Delta\phi_i(z)|$, $i = 1,\ldots,n$ and $\epsilon_{B} = \|B_n - \hat{B}_n\| = \|\Delta B_n\|$ will appear, in which $\hat{\Lambda}_n = \diag{\hat{\lambda}_1,\ldots,\hat{\lambda}_n}$. Therein, $\hat{\phi}_i(z)$ results from a suitable interpolation of \eqref{phiappm}. Small Krylov-DMD errors are indicated by a small equation error $r_ne_i\t v_i$, $i = 1,\ldots,n$, in \eqref{Kdmdm}, which can be evaluated based on data. While error bounds for the data-based approximation of the Koopman operator for nonlinear finite-dimensional systems are already available in the literature (see \cite{Nue23,Ph24}), it will be verified for the parabolic systems in question that the data-driven closed-loop system remains exponentially stable in the $L_2$-norm if the Krylov-DMD errors are sufficiently small. This is the result of the next theorem.
\begin{thm}[Data-driven closed-loop system]\label{thm:ddcl}
Assume that $\hat{\Lambda}_n - \hat{B}_nK$ is a Hurwitz matrix so that there exists the positive definite solution $\Pi$ of $ (\hat{\Lambda}_n - \hat{B}_nK)\t \Pi + \Pi(\hat{\Lambda}_n - \hat{B}_nK) = -I$ with $\lambda_{\text{max}}(\Pi)$ and $\lambda_{\text{min}}(\Pi)$ denoting its largest and smallest eigenvalue. Define $\underline{\hat{\varphi}}_n[x] = \col{\hat{\varphi}_1[x],\ldots,\hat{\varphi}_n[x]}$ and let $c_\phi > 0$ be such that $\|\uphi - \hat{\underline{\varphi}}_n[x]\|_{\mathbb{C}^n} \leq c_\phi\|\uphi\|_{\mathbb{C}^n}$, $x \in X$, $i = 1,\ldots,n$. Assume that $\gamma = 2((\epsilon_{\lambda} + \epsilon_B\|K\|)\lambda_{\text{max}}(\Pi) + c_\phi\|\Pi B_nK\|)-1$ is such that $\lambda_i < \hat{\alpha} < 0$, $i > n$, with $\hat{\alpha} = \gamma/(2\lambda_{\text{max}}(\Pi))$. Then, the closed-loop system \eqref{varphisys} with the state feedback $u(t) = -K\underline{\hat{\varphi}}_n[x(t)]$ is exponentially stable in the $L_2$-norm $\|\cdot\|$, i.e., $\|x(t)\| \leq \hat{M}_x\e^{\hat{\alpha} t}\|x(0)\|$, $t \geq 0$, $\forall x(0) \in X$
and an $\hat{M}_x \geq 1$ (cf. Theorem \ref{lem:stabl}).
\end{thm}
\begin{proof}
In view of \eqref{varphisys} it is readily verified that the data-driven closed-loop system takes the form
\begin{subequations}
\begin{align}
\dot{\underline{\varphi}}_n[x] &= (\hat{\Lambda}_n - \hat{B}_nK)\underline{\varphi}_n[x] 
                                  + (\Delta\Lambda_n -\Delta B_nK)\underline{\varphi}_n[x]\nonumber\\
                               & \quad  + B_nK\Delta\underline{\varphi}_n[x]\label{phinsys}\\
\dot{\varphi}_i[x] &= \lambda_i\varphi_i[x] - b_i\t K(\underline{\varphi}_n[x] - \Delta\underline{\varphi}_n[x]), \quad i > n,\label{phiisys}
\end{align}	 
\end{subequations}
with $\Delta\uphi = \uphi - \hat{\underline{\varphi}}_n[x]$. Consider the Lyapunov function candidate $V(\uphi) = \uphit \Pi\uphi$, $\Pi > 0$. Taking the Lyapunov equation in Theorem \ref{thm:ddcl} into account the result $\dot{V}(\uphi) = - \uphit\uphi + 2\uphit(\Pi(\Delta\Lambda_n-\Delta B_nK)\uphi + \Pi B_nK\Delta\uphi)$
holds on the solutions of \eqref{phinsys}. Note that $\|\Delta\uphi\|_{\mathbb{C}^n} \leq c_\phi\|\uphi\|_{\mathbb{C}^n}$, $x \in X$, follows from the estimate of Theorem \ref{thm:ddcl}, because $\Delta\uphi = \langle x,\col{\Delta\phi_1,\ldots,\Delta\phi_n}\rangle$. Then, it is readily verified that $\dot{V}(\uphi) \leq \gamma\|\uphi\|_{\mathbb{C}^n}^2$. The inequality $\lambda_{\text{min}}(\Pi)\|\uphi\|_{\mathbb{C}^n}^2 \leq V(\uphi) \leq \lambda_{\text{max}}(\Pi)\|\uphi\|^2_{\mathbb{C}^n}$ leads to $\dot{V}(\uphi) \leq (\gamma/\lambda_{\text{max}}(\Pi))V(\uphi)$. Hence, by making use of the \emph{comparison principle} the result $V(\underline{\varphi}_n[x(t)]) \leq \e^{(\gamma/\lambda_{\text{max}}(\Pi))t}V(\underline{\varphi}_n[x(0)])$, $t \geq 0$, follows (see, e.g., \cite[Ch. 3.4]{Khl02}). Then, one finds
$\|\underline{\varphi}_n[x(t)]\|_{\mathbb{C}^n} \leq \sqrt{\lambda_{\text{max}}(\Pi)/\lambda_{\text{min}}(\Pi)}\exp{\frac{\gamma}{2\lambda_{\text{max}}(\Pi)}t} \|\underline{\varphi}_n[x(0)]\|_{\mathbb{C}^n}$, $t \geq 0$, after a simple calculation, which shows that \eqref{phinsys} is exponentially stable under the conditions of Theorem \ref{thm:ddcl}. Using the estimates of the latter theorem, the solution of \eqref{phiisys} can be estimated by $|\varphi_i[x(t)]|^2 \leq 2\e^{2\lambda_it}|\varphi_i[x(0)]|^2  + 2\int_0^t\e^{2\lambda_i(t-\tau)}\|b_i\|^2_{\mathbb{C}^n} \|K\|(1+c_\phi)^2\|\underline{\varphi}_n[x(\tau)]\|_{\mathbb{C}^n}^2\d\tau$ for $t \geq 0$ and $i > n$. With this, the remaining part of the proof follows the same lines as the proof of Theorem \ref{lem:stabl}. 
\end{proof}		
\begin{rem}\label{rem:K}	
Note that the condition $\epsilon_i(z) \leq c_\phi|\phi_i(z)|$, $i = 1,\ldots,n$, implies $\|\uphi - \hat{\underline{\varphi}}_n[x]\|_{\mathbb{C}^n} \leq c_\phi\|\uphi\|_{\mathbb{C}^n}$, $x \in X$, $i = 1,\ldots,n$. Hence, small Krylov errors $\epsilon_i(z)$ lead to a small $c_\phi$.
 \hfill $\triangleleft$	 
\end{rem}

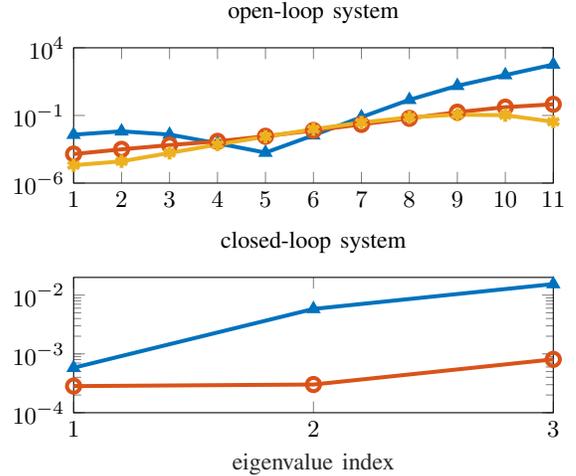
\begin{figure}[t!]\centering	
	\begin{tabular}{l}
		\input{plant_error_2.tex}
		\\
		\input{control_error.tex}
	\end{tabular}
	\caption{Plots of the eigenvalue error (\ref{m1}), the Koopman mode error (\ref{m2}) and the relative residual (\ref{m3}) for the open-loop  system in the first row. The second row shows the errors for the closed-loop eigenvalue (\ref{m4}) and the related eigenvectors $\hat{\tilde{\psi}}_i$ (\ref{m5}) of $\tilde{\mathcal{A}}^*$ defining the assigned closed-loop eigenfunctionals \eqref{clefktl}.}\label{fig:ex1}%
\end{figure}	

\section{Example}\label{sec:ex}
\textbf{System and State Data.} The results of the paper are illustrated for an unstable system \eqref{plant} with $a(z) = (7-8(z - 0.5)^2)$, $q_0 = 2$ and $q_1 = 1$. Thereby, $\rho = 1$ is assumed and known so that only the state data matrix \eqref{xdata} is required in the data-driven system analysis, which results from a simulation in Matlab.
In order to obtain good data for the system, a suitable IC has to be prepared. This is achieved by applying a rectangular impulse $u_1(t) = 10(s(t + 0.1) - s(t))$, $u_2 \equiv 0$ and $x(z,-0.1) = 0$, in which $s(t)$ is the step function. The sampling time $t_s$ is determined by successively increasing $t_s$ and selecting that $n$, for which the norm $\|r_n\|$ is sufficiently small. With this, the number $n$ of essential modes contained in the data for a particular $t_s$ can be estimated (cf. Remark \ref{rem:nchoose}). For the considered setup the sampling time $t_s =  0.004$ allows to get $n = 11$ Koopman modes and eigenvalues with $\|r_n\| = 1.5886\cdot10^{-7}$ in \eqref{lkf}. Then, in a first step the output data $\mathbbm{D}_{12}(x) \in \mathbb{R}^{500 \times 12}$ for $M = 500$ spatial sampling points is obtained from the simulation by assuming $\epsilon$ in \eqref{cdef} sufficiently small, which means that \emph{state data} is considered.

\textbf{Krylov-DMD Analysis of the Plant.} With $n$ determined, the coefficient vector $f \in \mathbb{R}^{11}$ follows from \eqref{fdet}. Then, the eigenvalues $\mu_i$, $i = 1,\ldots,11$ and the corresponding eigenvectors $v_i$ of $F$ are calculated by solving the eigenvalue problem for the companion matrix $F$ using \texttt{eig} in Matlab. With this, the sampled Koopman modes are obtained from \eqref{phiappm} and the corresponding Koopman eigenvalues are $\hat{\lambda}_i = \ln\mu_i/t_s$. The accuracy of the Krylov-DMD is shown in the first row of Figure \ref{fig:ex1}. For this, the eigenvalue error $|\hat{\lambda}_i - \lambda_i|$, the Koopman mode error $\|\hat{\phi}_i - \phi_i\|$ and the relative residual $(\|r_n\|_{\mathbb{R}^M}|e_n\t v_i|)/\|\mathbbm{D}_{n}(x)v_i\|_{\mathbb{R}^M}$ is plotted (cf. \eqref{Kdmdm}). Note that due to the fine spatial sampling with $M = 500$ the result $\hat{\phi}_i$ is treated as a function so that the spatial samples can be utilized to directly approximate the $L_2$-error norm. As expected, the first seven Koopman modes are obtained with a small error, which is also indicated by a small relative residual. This is in accordance with a spectral analysis of $x(0)$, which shows that only the first seven are dominant in the data. Thereby, the relative residual, which is obtained from data, qualifies as a good indicator for the Koopman mode and eigenvalue accuracy. In order to be able to use \emph{measurement data} so-called \emph{delay coordinates} (see, e.g., \cite{Ka20}) can be employed and the Krylov-DMD is applied to the resulting data matrix. Then, the same result is obtained by using $M = 5$ pointlike measurements in \eqref{xdata} with three delay coordinates. 

\begin{figure}[t!]\centering	
	\begin{tabular}{l}
		\input{control_state.tex}
	\end{tabular}
	\caption{Solution $x(z,t)$ of the closed-loop system for the initial condition used to collect the data.}\label{fig:ex2}%
\end{figure}
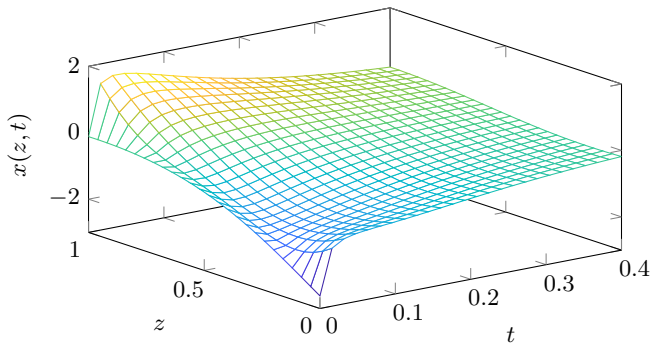

\textbf{Data-Driven Koopman Eigenstructure Assignment.} In order to stabilize the system, the eigenvalue assignment $\sigma(\hat{\Lambda}_n - \hat{B}_nK) = \{-\hat{\lambda}_1,2\hat{\lambda}_2, 1.5\hat{\lambda}_3\} = \{-7.0034, -10.771, -52.729\}$, where $\hat{\lambda}_i$, $i = 1,2,3$, are the obtained open-loop eigenvalues by applying  the Krylov-DMD including the unstable eigenvalue $\hat{\lambda}_1$. The remaining degrees of freedom contained in $K$ after eigenvalue assignment are contained in the parameter vectors $p_i \in \mathbb{R}^2$, $i = 1,2,3$. They are calculated with \texttt{fmincon} to minimize $\|K\|$ in Matlab by constraining the elements of $p_i$ in $[-1,1]$ in view of the condition for $\gamma$ in Theorem \ref{thm:ddcl}. This yields the feedback gain $K$ with $\|K\| = 18.814$ assigning the specified closed-loop spectrum $\sigma(\hat{\Lambda}_n - \hat{B}_nK)$ and the closed-loop eigenfunctionals $\hat{\tilde{\varphi}}_i[x] = \langle x,\hat{\tilde{\psi}}_i\rangle$, which can be calculated with \eqref{tilphipar}. 
The accuracy of the Koopman eigenstructure assignment is shown in the second row of Figure \ref{fig:ex1}. Obviously, the resulting closed-loop eigenvalues are subject to only small errors. Thereby, the remaining dynamics is only marginally shifted, which was verified numerically. Similarly, the $L_2$-error for the assigned $\hat{\tilde{\psi}}_i$, $i = 1,2,3$, is also small. The simulation of the closed-loop system for $x(0)$ used to collect the data is based on a FEM model using $2001$ grid points. Figure \ref{fig:ex2} depicts the resulting solution, which verifies closed-loop stability with a desired performance. A faster closed-loop systems leads to larger errors for the assigned closed-loop eigenvalues and -functionals. In this case, the SVD-enhanced Krylov-DMD methods presented in \cite{Ar17} yield superior results so that they are better suited for a fast closed-loop dynamics assignment.%

\section{Concluding Remarks}
The results of the paper can directly be extended to other distributed-parameter systems amenable to a modal approach. This includes biharmonic PDEs (beam models) and classes of the hyperbolic PDEs (wave equations). 
Since for distributed-parameter systems the state is not available for feedback, observers have to be designed to determine output feedback controllers. They can also be based on a Koopman modal approach, which will be investigated in future work. Finally, it is of paramount interest to extend the results of the paper to nonlinear PDEs. 

\section*{Acknowledgment}
The author would like to thank one of the anonymous reviewers for helpful comments.



\printbibliography

\end{document}

%% file: plant_error_2.tex
%
%
\definecolor{mycolor1}{rgb}{0.00000,0.44700,0.74100}%
\definecolor{mycolor2}{rgb}{0.85000,0.32500,0.09800}%
\definecolor{mycolor3}{rgb}{0.92900,0.69400,0.12500}%
\begin{tikzpicture}

\begin{axis}[%
scale=0.9,
width=0.8\linewidth,
height=2cm,
at={(0,0)},
scale only axis,
xmin=1,
xmax=11,
xtick={1, 2,3,4,5,6,7,8,9,10,11},
ymode=log,
yminorticks=true,
ymin=1e-06,
ymax=10000,
axis background/.style={fill=white},
legend style={legend cell align=left, align=left, draw=white!15!black, at={(-0.,1)}, anchor=north west},
title=open-loop system
]
\addplot [color=mycolor1, line width=1.5pt, mark size=1.7pt, mark=triangle, mark options={solid, mycolor1}]
  table[row sep=crcr]{%
1	0.00379162300070313\\
2	0.00672842940423024\\
3	0.00386853558602951\\
4	0.000871866152465373\\
5	0.000178723883465182\\
6	0.00341437657206711\\
7	0.0751889829780907\\
8	1.43048552534589\\
9	15.5448138742169\\
10	96.6022014788371\\
11	581.025374223971\\
};\label{m1}

\addplot [color=mycolor2, line width=1.5pt, mark size=2.5pt, mark=o, mark options={solid, mycolor2}]
  table[row sep=crcr]{%
1	0.00014018649642484\\
2	0.000312743462935884\\
3	0.000644557236619712\\
4	0.00122074151136572\\
5	0.00289388016366846\\
6	0.00778831103533153\\
7	0.0218384327933904\\
8	0.0615264238723342\\
9	0.174611087984298\\
10	0.409168434463695\\
11	0.669424465624492\\
};\label{m2}

\addplot [color=mycolor3, line width=1.5pt, mark size=2.5pt, mark=asterisk, mark options={solid, mycolor3}]
  table[row sep=crcr]{%
1	2.09237451985609e-05\\
2	4.06071454905625e-05\\
3	0.000168679523886102\\
4	0.000705761524751859\\
5	0.00267915883406907\\
6	0.0095509826640145\\
7	0.0301260778328917\\
8	0.0730297787879556\\
9	0.114175123183448\\
10	0.104190375465415\\
11	0.0341684538055019\\
};\label{m3}
\end{axis}

\begin{axis}[%
scale=0.9,
width=0.8\linewidth,
height=2cm,
at={(0in,0in)},
scale only axis,
xmin=0,
xmax=1,
ymin=0,
ymax=1,
axis line style={draw=none},
ticks=none,
axis x line*=bottom,
axis y line*=left
]
\end{axis}
\end{tikzpicture}%

%% file: control_error.tex
%
%
\definecolor{mycolor1}{rgb}{0.00000,0.44700,0.74100}%
\definecolor{mycolor2}{rgb}{0.85000,0.32500,0.09800}%


%
\begin{tikzpicture}

\begin{axis}[%
scale=0.9,
width=0.8\linewidth,
height=2cm,
at={(0,0)},
scale only axis,
xmin=1,
xmax=3,
xtick={1,2,3},
xlabel style={font=\color{white!15!black}},
xlabel={eigenvalue index},
ymode=log,
ymin=0.0001,
ymax=0.02,
axis background/.style={fill=white},
legend style={legend cell align=left, align=left, draw=white!15!black, at={(-0.,1)}, anchor=north west},
title=closed-loop system
]
\addplot [color=mycolor1, line width=1.5pt, mark size=1.7pt, mark=triangle, mark options={solid, mycolor1}]
  table[row sep=crcr]{%
1	0.000581005621015152\\
2	0.00581883080049828\\
3	0.0154085249008773\\
}; \label{m4}

\addplot [color=mycolor2, line width=1.5pt, mark size=2.5pt, mark=o, mark options={solid, mycolor2}]
  table[row sep=crcr]{%
1	0.000283454700925472\\
2	0.000301589338885969\\
3	0.00080798443502036\\
}; \label{m5}

\end{axis}

\begin{axis}[%
scale=0.9,
width=0.8\linewidth,
height=2cm,
at={(0,0)},
scale only axis,
xmin=0,
xmax=1,
ymin=0,
ymax=1,
axis line style={draw=none},
ticks=none,
axis x line*=bottom,
axis y line*=left
]
\end{axis}
\end{tikzpicture}%

%% file: control_state.tex
%
%
\begin{tikzpicture}

\begin{axis}[%
width=0.8\linewidth,
height=4cm,
at={(0,0)},
scale only axis,
xmin=0,
xmax=0.4,
xlabel={$t$},
ymin=0,
ymax=1,
ylabel={$z$},
zmin=-3,
zmax=2,
zlabel={$x(z,t)$},
view={-37.5}{30},
]

\addplot3[%
surf,
shader=flat corner, fill=white, z buffer=sort, colormap={mymap}{[1pt] rgb(0pt)=(0.2422,0.1504,0.6603); rgb(1pt)=(0.2444,0.1534,0.6728); rgb(2pt)=(0.2464,0.1569,0.6847); rgb(3pt)=(0.2484,0.1607,0.6961); rgb(4pt)=(0.2503,0.1648,0.7071); rgb(5pt)=(0.2522,0.1689,0.7179); rgb(6pt)=(0.254,0.1732,0.7286); rgb(7pt)=(0.2558,0.1773,0.7393); rgb(8pt)=(0.2576,0.1814,0.7501); rgb(9pt)=(0.2594,0.1854,0.761); rgb(11pt)=(0.2628,0.1932,0.7828); rgb(12pt)=(0.2645,0.1972,0.7937); rgb(13pt)=(0.2661,0.2011,0.8043); rgb(14pt)=(0.2676,0.2052,0.8148); rgb(15pt)=(0.2691,0.2094,0.8249); rgb(16pt)=(0.2704,0.2138,0.8346); rgb(17pt)=(0.2717,0.2184,0.8439); rgb(18pt)=(0.2729,0.2231,0.8528); rgb(19pt)=(0.274,0.228,0.8612); rgb(20pt)=(0.2749,0.233,0.8692); rgb(21pt)=(0.2758,0.2382,0.8767); rgb(22pt)=(0.2766,0.2435,0.884); rgb(23pt)=(0.2774,0.2489,0.8908); rgb(24pt)=(0.2781,0.2543,0.8973); rgb(25pt)=(0.2788,0.2598,0.9035); rgb(26pt)=(0.2794,0.2653,0.9094); rgb(27pt)=(0.2798,0.2708,0.915); rgb(28pt)=(0.2802,0.2764,0.9204); rgb(29pt)=(0.2806,0.2819,0.9255); rgb(30pt)=(0.2809,0.2875,0.9305); rgb(31pt)=(0.2811,0.293,0.9352); rgb(32pt)=(0.2813,0.2985,0.9397); rgb(33pt)=(0.2814,0.304,0.9441); rgb(34pt)=(0.2814,0.3095,0.9483); rgb(35pt)=(0.2813,0.315,0.9524); rgb(36pt)=(0.2811,0.3204,0.9563); rgb(37pt)=(0.2809,0.3259,0.96); rgb(38pt)=(0.2807,0.3313,0.9636); rgb(39pt)=(0.2803,0.3367,0.967); rgb(40pt)=(0.2798,0.3421,0.9702); rgb(41pt)=(0.2791,0.3475,0.9733); rgb(42pt)=(0.2784,0.3529,0.9763); rgb(43pt)=(0.2776,0.3583,0.9791); rgb(44pt)=(0.2766,0.3638,0.9817); rgb(45pt)=(0.2754,0.3693,0.984); rgb(46pt)=(0.2741,0.3748,0.9862); rgb(47pt)=(0.2726,0.3804,0.9881); rgb(48pt)=(0.271,0.386,0.9898); rgb(49pt)=(0.2691,0.3916,0.9912); rgb(50pt)=(0.267,0.3973,0.9924); rgb(51pt)=(0.2647,0.403,0.9935); rgb(52pt)=(0.2621,0.4088,0.9946); rgb(53pt)=(0.2591,0.4145,0.9955); rgb(54pt)=(0.2556,0.4203,0.9965); rgb(55pt)=(0.2517,0.4261,0.9974); rgb(56pt)=(0.2473,0.4319,0.9983); rgb(57pt)=(0.2424,0.4378,0.9991); rgb(58pt)=(0.2369,0.4437,0.9996); rgb(59pt)=(0.2311,0.4497,0.9995); rgb(60pt)=(0.225,0.4559,0.9985); rgb(61pt)=(0.2189,0.462,0.9968); rgb(62pt)=(0.2128,0.4682,0.9948); rgb(63pt)=(0.2066,0.4743,0.9926); rgb(64pt)=(0.2006,0.4803,0.9906); rgb(65pt)=(0.195,0.4861,0.9887); rgb(66pt)=(0.1903,0.4919,0.9867); rgb(67pt)=(0.1869,0.4975,0.9844); rgb(68pt)=(0.1847,0.503,0.9819); rgb(69pt)=(0.1831,0.5084,0.9793); rgb(70pt)=(0.1818,0.5138,0.9766); rgb(71pt)=(0.1806,0.5191,0.9738); rgb(72pt)=(0.1795,0.5244,0.9709); rgb(73pt)=(0.1785,0.5296,0.9677); rgb(74pt)=(0.1778,0.5349,0.9641); rgb(75pt)=(0.1773,0.5401,0.9602); rgb(76pt)=(0.1768,0.5452,0.956); rgb(77pt)=(0.1764,0.5504,0.9516); rgb(78pt)=(0.1755,0.5554,0.9473); rgb(79pt)=(0.174,0.5605,0.9432); rgb(80pt)=(0.1716,0.5655,0.9393); rgb(81pt)=(0.1686,0.5705,0.9357); rgb(82pt)=(0.1649,0.5755,0.9323); rgb(83pt)=(0.161,0.5805,0.9289); rgb(84pt)=(0.1573,0.5854,0.9254); rgb(85pt)=(0.154,0.5902,0.9218); rgb(86pt)=(0.1513,0.595,0.9182); rgb(87pt)=(0.1492,0.5997,0.9147); rgb(88pt)=(0.1475,0.6043,0.9113); rgb(89pt)=(0.1461,0.6089,0.908); rgb(90pt)=(0.1446,0.6135,0.905); rgb(91pt)=(0.1429,0.618,0.9022); rgb(92pt)=(0.1408,0.6226,0.8998); rgb(93pt)=(0.1383,0.6272,0.8975); rgb(94pt)=(0.1354,0.6317,0.8953); rgb(95pt)=(0.1321,0.6363,0.8932); rgb(96pt)=(0.1288,0.6408,0.891); rgb(97pt)=(0.1253,0.6453,0.8887); rgb(98pt)=(0.1219,0.6497,0.8862); rgb(99pt)=(0.1185,0.6541,0.8834); rgb(100pt)=(0.1152,0.6584,0.8804); rgb(101pt)=(0.1119,0.6627,0.877); rgb(102pt)=(0.1085,0.6669,0.8734); rgb(103pt)=(0.1048,0.671,0.8695); rgb(104pt)=(0.1009,0.675,0.8653); rgb(105pt)=(0.0964,0.6789,0.8609); rgb(106pt)=(0.0914,0.6828,0.8562); rgb(107pt)=(0.0855,0.6865,0.8513); rgb(108pt)=(0.0789,0.6902,0.8462); rgb(109pt)=(0.0713,0.6938,0.8409); rgb(110pt)=(0.0628,0.6972,0.8355); rgb(111pt)=(0.0535,0.7006,0.8299); rgb(112pt)=(0.0433,0.7039,0.8242); rgb(113pt)=(0.0328,0.7071,0.8183); rgb(114pt)=(0.0234,0.7103,0.8124); rgb(115pt)=(0.0155,0.7133,0.8064); rgb(116pt)=(0.0091,0.7163,0.8003); rgb(117pt)=(0.0046,0.7192,0.7941); rgb(118pt)=(0.0019,0.722,0.7878); rgb(119pt)=(0.0009,0.7248,0.7815); rgb(120pt)=(0.0018,0.7275,0.7752); rgb(121pt)=(0.0046,0.7301,0.7688); rgb(122pt)=(0.0094,0.7327,0.7623); rgb(123pt)=(0.0162,0.7352,0.7558); rgb(124pt)=(0.0253,0.7376,0.7492); rgb(125pt)=(0.0369,0.74,0.7426); rgb(126pt)=(0.0504,0.7423,0.7359); rgb(127pt)=(0.0638,0.7446,0.7292); rgb(128pt)=(0.077,0.7468,0.7224); rgb(129pt)=(0.0899,0.7489,0.7156); rgb(130pt)=(0.1023,0.751,0.7088); rgb(131pt)=(0.1141,0.7531,0.7019); rgb(132pt)=(0.1252,0.7552,0.695); rgb(133pt)=(0.1354,0.7572,0.6881); rgb(134pt)=(0.1448,0.7593,0.6812); rgb(135pt)=(0.1532,0.7614,0.6741); rgb(136pt)=(0.1609,0.7635,0.6671); rgb(137pt)=(0.1678,0.7656,0.6599); rgb(138pt)=(0.1741,0.7678,0.6527); rgb(139pt)=(0.1799,0.7699,0.6454); rgb(140pt)=(0.1853,0.7721,0.6379); rgb(141pt)=(0.1905,0.7743,0.6303); rgb(142pt)=(0.1954,0.7765,0.6225); rgb(143pt)=(0.2003,0.7787,0.6146); rgb(144pt)=(0.2061,0.7808,0.6065); rgb(145pt)=(0.2118,0.7828,0.5983); rgb(146pt)=(0.2178,0.7849,0.5899); rgb(147pt)=(0.2244,0.7869,0.5813); rgb(148pt)=(0.2318,0.7887,0.5725); rgb(149pt)=(0.2401,0.7905,0.5636); rgb(150pt)=(0.2491,0.7922,0.5546); rgb(151pt)=(0.2589,0.7937,0.5454); rgb(152pt)=(0.2695,0.7951,0.536); rgb(153pt)=(0.2809,0.7964,0.5266); rgb(154pt)=(0.2929,0.7975,0.517); rgb(155pt)=(0.3052,0.7985,0.5074); rgb(156pt)=(0.3176,0.7994,0.4975); rgb(157pt)=(0.3301,0.8002,0.4876); rgb(158pt)=(0.3424,0.8009,0.4774); rgb(159pt)=(0.3548,0.8016,0.4669); rgb(160pt)=(0.3671,0.8021,0.4563); rgb(161pt)=(0.3795,0.8026,0.4454); rgb(162pt)=(0.3921,0.8029,0.4344); rgb(163pt)=(0.405,0.8031,0.4233); rgb(164pt)=(0.4184,0.803,0.4122); rgb(165pt)=(0.4322,0.8028,0.4013); rgb(166pt)=(0.4463,0.8024,0.3904); rgb(167pt)=(0.4608,0.8018,0.3797); rgb(168pt)=(0.4753,0.8011,0.3691); rgb(169pt)=(0.4899,0.8002,0.3586); rgb(170pt)=(0.5044,0.7993,0.348); rgb(171pt)=(0.5187,0.7982,0.3374); rgb(172pt)=(0.5329,0.797,0.3267); rgb(173pt)=(0.547,0.7957,0.3159); rgb(175pt)=(0.5748,0.7929,0.2941); rgb(176pt)=(0.5886,0.7913,0.2833); rgb(177pt)=(0.6024,0.7896,0.2726); rgb(178pt)=(0.6161,0.7878,0.2622); rgb(179pt)=(0.6297,0.7859,0.2521); rgb(180pt)=(0.6433,0.7839,0.2423); rgb(181pt)=(0.6567,0.7818,0.2329); rgb(182pt)=(0.6701,0.7796,0.2239); rgb(183pt)=(0.6833,0.7773,0.2155); rgb(184pt)=(0.6963,0.775,0.2075); rgb(185pt)=(0.7091,0.7727,0.1998); rgb(186pt)=(0.7218,0.7703,0.1924); rgb(187pt)=(0.7344,0.7679,0.1852); rgb(188pt)=(0.7468,0.7654,0.1782); rgb(189pt)=(0.759,0.7629,0.1717); rgb(190pt)=(0.771,0.7604,0.1658); rgb(191pt)=(0.7829,0.7579,0.1608); rgb(192pt)=(0.7945,0.7554,0.157); rgb(193pt)=(0.806,0.7529,0.1546); rgb(194pt)=(0.8172,0.7505,0.1535); rgb(195pt)=(0.8281,0.7481,0.1536); rgb(196pt)=(0.8389,0.7457,0.1546); rgb(197pt)=(0.8495,0.7435,0.1564); rgb(198pt)=(0.86,0.7413,0.1587); rgb(199pt)=(0.8703,0.7392,0.1615); rgb(200pt)=(0.8804,0.7372,0.165); rgb(201pt)=(0.8903,0.7353,0.1695); rgb(202pt)=(0.9,0.7336,0.1749); rgb(203pt)=(0.9093,0.7321,0.1815); rgb(204pt)=(0.9184,0.7308,0.189); rgb(205pt)=(0.9272,0.7298,0.1973); rgb(206pt)=(0.9357,0.729,0.2061); rgb(207pt)=(0.944,0.7285,0.2151); rgb(208pt)=(0.9523,0.7284,0.2237); rgb(209pt)=(0.9606,0.7285,0.2312); rgb(210pt)=(0.9689,0.7292,0.2373); rgb(211pt)=(0.977,0.7304,0.2418); rgb(212pt)=(0.9842,0.733,0.2446); rgb(213pt)=(0.99,0.7365,0.2429); rgb(214pt)=(0.9946,0.7407,0.2394); rgb(215pt)=(0.9966,0.7458,0.2351); rgb(216pt)=(0.9971,0.7513,0.2309); rgb(217pt)=(0.9972,0.7569,0.2267); rgb(218pt)=(0.9971,0.7626,0.2224); rgb(219pt)=(0.9969,0.7683,0.2181); rgb(220pt)=(0.9966,0.774,0.2138); rgb(221pt)=(0.9962,0.7798,0.2095); rgb(222pt)=(0.9957,0.7856,0.2053); rgb(223pt)=(0.9949,0.7915,0.2012); rgb(224pt)=(0.9938,0.7974,0.1974); rgb(225pt)=(0.9923,0.8034,0.1939); rgb(226pt)=(0.9906,0.8095,0.1906); rgb(227pt)=(0.9885,0.8156,0.1875); rgb(228pt)=(0.9861,0.8218,0.1846); rgb(229pt)=(0.9835,0.828,0.1817); rgb(230pt)=(0.9807,0.8342,0.1787); rgb(231pt)=(0.9778,0.8404,0.1757); rgb(232pt)=(0.9748,0.8467,0.1726); rgb(233pt)=(0.972,0.8529,0.1695); rgb(234pt)=(0.9694,0.8591,0.1665); rgb(235pt)=(0.9671,0.8654,0.1636); rgb(236pt)=(0.9651,0.8716,0.1608); rgb(237pt)=(0.9634,0.8778,0.1582); rgb(238pt)=(0.9619,0.884,0.1557); rgb(239pt)=(0.9608,0.8902,0.1532); rgb(240pt)=(0.9601,0.8963,0.1507); rgb(241pt)=(0.9596,0.9023,0.148); rgb(242pt)=(0.9595,0.9084,0.145); rgb(243pt)=(0.9597,0.9143,0.1418); rgb(244pt)=(0.9601,0.9203,0.1382); rgb(245pt)=(0.9608,0.9262,0.1344); rgb(246pt)=(0.9618,0.932,0.1304); rgb(247pt)=(0.9629,0.9379,0.1261); rgb(248pt)=(0.9642,0.9437,0.1216); rgb(249pt)=(0.9657,0.9494,0.1168); rgb(250pt)=(0.9674,0.9552,0.1116); rgb(251pt)=(0.9692,0.9609,0.1061); rgb(252pt)=(0.9711,0.9667,0.1001); rgb(253pt)=(0.973,0.9724,0.0938); rgb(254pt)=(0.9749,0.9782,0.0872); rgb(255pt)=(0.9769,0.9839,0.0805)}, mesh/rows=26]
table[row sep=crcr, point meta=\thisrow{c}] {%
x	y	z	c\\
0	0	-2.62103366464545	-2.62103366464545\\
0.016	0	-1.25629488286272	-1.25629488286272\\
0.032	0	-1.04193398965811	-1.04193398965811\\
0.048	0	-0.954837210574449	-0.954837210574449\\
0.064	0	-0.908150818052292	-0.908150818052292\\
0.08	0	-0.872996691286192	-0.872996691286192\\
0.096	0	-0.838596303382896	-0.838596303382896\\
0.112	0	-0.801557691721352	-0.801557691721352\\
0.128	0	-0.7614898949289	-0.7614898949289\\
0.144	0	-0.719089635211603	-0.719089635211603\\
0.16	0	-0.675332788332385	-0.675332788332385\\
0.176	0	-0.631155813978112	-0.631155813978112\\
0.192	0	-0.587347911951497	-0.587347911951497\\
0.208	0	-0.544530004272502	-0.544530004272502\\
0.224	0	-0.503166863678906	-0.503166863678906\\
0.24	0	-0.463589745288879	-0.463589745288879\\
0.256	0	-0.426020356963167	-0.426020356963167\\
0.272	0	-0.390592758674592	-0.390592758674592\\
0.288	0	-0.35737218116815	-0.35737218116815\\
0.304	0	-0.326370713244196	-0.326370713244196\\
0.32	0	-0.297560155632539	-0.297560155632539\\
0.336	0	-0.270882435345241	-0.270882435345241\\
0.352	0	-0.246257970377925	-0.246257970377925\\
0.368	0	-0.223592335096214	-0.223592335096214\\
0.384	0	-0.202781528121648	-0.202781528121648\\
0.4	0	-0.183716097410287	-0.183716097410287\\
0	0.04	-2.43035917923046	-2.43035917923046\\
0.016	0.04	-1.43656917723453	-1.43656917723453\\
0.032	0.04	-1.17963738447827	-1.17963738447827\\
0.048	0.04	-1.06835889233913	-1.06835889233913\\
0.064	0.04	-1.00637884269903	-1.00637884269903\\
0.08	0.04	-0.960174538612636	-0.960174538612636\\
0.096	0.04	-0.916906111864418	-0.916906111864418\\
0.112	0.04	-0.872271885323106	-0.872271885323106\\
0.128	0.04	-0.82546992622692	-0.82546992622692\\
0.144	0.04	-0.77700055919813	-0.77700055919813\\
0.16	0.04	-0.727735552117449	-0.727735552117449\\
0.176	0.04	-0.678547213164357	-0.678547213164357\\
0.192	0.04	-0.630178283173374	-0.630178283173374\\
0.208	0.04	-0.583211529484411	-0.583211529484411\\
0.224	0.04	-0.538077968983092	-0.538077968983092\\
0.24	0.04	-0.495077906614305	-0.495077906614305\\
0.256	0.04	-0.454404270551652	-0.454404270551652\\
0.272	0.04	-0.416164259852131	-0.416164259852131\\
0.288	0.04	-0.380398054447477	-0.380398054447477\\
0.304	0.04	-0.347094438807313	-0.347094438807313\\
0.32	0.04	-0.316203599455306	-0.316203599455306\\
0.336	0.04	-0.287647477463464	-0.287647477463464\\
0.352	0.04	-0.261328063076419	-0.261328063076419\\
0.368	0.04	-0.23713398381661	-0.23713398381661\\
0.384	0.04	-0.214945690024942	-0.214945690024942\\
0.4	0.04	-0.194639494765222	-0.194639494765222\\
0	0.08	-2.24015762079182	-2.24015762079182\\
0.016	0.08	-1.5641626134231	-1.5641626134231\\
0.032	0.08	-1.29166925750249	-1.29166925750249\\
0.048	0.08	-1.16503516500557	-1.16503516500557\\
0.064	0.08	-1.09099976535444	-1.09099976535444\\
0.08	0.08	-1.03485508581493	-1.03485508581493\\
0.096	0.08	-0.98307467612246	-0.98307467612246\\
0.112	0.08	-0.931013064574098	-0.931013064574098\\
0.128	0.08	-0.877672163741377	-0.877672163741377\\
0.144	0.08	-0.823421108294078	-0.823421108294078\\
0.16	0.08	-0.769032626211616	-0.769032626211616\\
0.176	0.08	-0.715298483309889	-0.715298483309889\\
0.192	0.08	-0.662893204142233	-0.662893204142233\\
0.208	0.08	-0.612340551256676	-0.612340551256676\\
0.224	0.08	-0.564019953723391	-0.564019953723391\\
0.24	0.08	-0.518186362330115	-0.518186362330115\\
0.256	0.08	-0.474992712198916	-0.474992712198916\\
0.272	0.08	-0.434510875836518	-0.434510875836518\\
0.288	0.08	-0.396749794294311	-0.396749794294311\\
0.304	0.08	-0.361670606160594	-0.361670606160594\\
0.32	0.08	-0.329199016621885	-0.329199016621885\\
0.336	0.08	-0.299235276198278	-0.299235276198278\\
0.352	0.08	-0.271662148042025	-0.271662148042025\\
0.368	0.08	-0.246351208721483	-0.246351208721483\\
0.384	0.08	-0.22316778117997	-0.22316778117997\\
0.4	0.08	-0.201974752358102	-0.201974752358102\\
0	0.12	-2.05220286738773	-2.05220286738773\\
0.016	0.12	-1.63665147635618	-1.63665147635618\\
0.032	0.12	-1.37478274376359	-1.37478274376359\\
0.048	0.12	-1.24218056504045	-1.24218056504045\\
0.064	0.12	-1.15972637981658	-1.15972637981658\\
0.08	0.12	-1.09502873207239	-1.09502873207239\\
0.096	0.12	-1.03530941107845	-1.03530941107845\\
0.112	0.12	-0.976171837593969	-0.976171837593969\\
0.128	0.12	-0.91664797017455	-0.91664797017455\\
0.144	0.12	-0.857046238625377	-0.857046238625377\\
0.16	0.12	-0.798048222090531	-0.798048222090531\\
0.176	0.12	-0.740350541155564	-0.740350541155564\\
0.192	0.12	-0.684539065642601	-0.684539065642601\\
0.208	0.12	-0.631058785013977	-0.631058785013977\\
0.224	0.12	-0.580220651134773	-0.580220651134773\\
0.24	0.12	-0.532220702710563	-0.532220702710563\\
0.256	0.12	-0.487161437890287	-0.487161437890287\\
0.272	0.12	-0.445071640220093	-0.445071640220093\\
0.288	0.12	-0.405923467251301	-0.405923467251301\\
0.304	0.12	-0.369646658596899	-0.369646658596899\\
0.32	0.12	-0.336140108510087	-0.336140108510087\\
0.336	0.12	-0.305281162681993	-0.305281162681993\\
0.352	0.12	-0.276933004475615	-0.276933004475615\\
0.368	0.12	-0.250950461687544	-0.250950461687544\\
0.384	0.12	-0.227184519863668	-0.227184519863668\\
0.4	0.12	-0.205485783538639	-0.205485783538639\\
0	0.16	-1.86821267086551	-1.86821267086551\\
0.016	0.16	-1.65651869487921	-1.65651869487921\\
0.032	0.16	-1.42740312268914	-1.42740312268914\\
0.048	0.16	-1.29781530790012	-1.29781530790012\\
0.064	0.16	-1.21060623495217	-1.21060623495217\\
0.08	0.16	-1.13888226493724	-1.13888226493724\\
0.096	0.16	-1.07197147577825	-1.07197147577825\\
0.112	0.16	-1.00628663080013	-1.00628663080013\\
0.128	0.16	-0.941102207734205	-0.941102207734205\\
0.144	0.16	-0.876731596708624	-0.876731596708624\\
0.16	0.16	-0.81377232910774	-0.81377232910774\\
0.176	0.16	-0.752812146206589	-0.752812146206589\\
0.192	0.16	-0.694329271727912	-0.694329271727912\\
0.208	0.16	-0.638671714072209	-0.638671714072209\\
0.224	0.16	-0.586066545702218	-0.586066545702218\\
0.24	0.16	-0.536638684615587	-0.536638684615587\\
0.256	0.16	-0.490430945777895	-0.490430945777895\\
0.272	0.16	-0.447422314010192	-0.447422314010192\\
0.288	0.16	-0.407543539863373	-0.407543539863373\\
0.304	0.16	-0.370690015547973	-0.370690015547973\\
0.32	0.16	-0.336732197454348	-0.336732197454348\\
0.336	0.16	-0.305523925918137	-0.305523925918137\\
0.352	0.16	-0.276908988201907	-0.276908988201907\\
0.368	0.16	-0.250726234615758	-0.250726234615758\\
0.384	0.16	-0.22681351388315	-0.22681351388315\\
0.4	0.16	-0.205010651483218	-0.205010651483218\\
0	0.2	-1.68979570493802	-1.68979570493802\\
0.016	0.2	-1.63010644829462	-1.63010644829462\\
0.032	0.2	-1.44958928279342	-1.44958928279342\\
0.048	0.2	-1.33070958316264	-1.33070958316264\\
0.064	0.2	-1.24207873258596	-1.24207873258596\\
0.08	0.2	-1.16485800247426	-1.16485800247426\\
0.096	0.2	-1.09163301018908	-1.09163301018908\\
0.112	0.2	-1.0200972986448	-1.0200972986448\\
0.128	0.2	-0.949942617987805	-0.949942617987805\\
0.144	0.2	-0.881538608631647	-0.881538608631647\\
0.16	0.2	-0.815401692579214	-0.815401692579214\\
0.176	0.2	-0.751997041832529	-0.751997041832529\\
0.192	0.2	-0.691677845682836	-0.691677845682836\\
0.208	0.2	-0.634678960643076	-0.634678960643076\\
0.224	0.2	-0.581130195373737	-0.581130195373737\\
0.24	0.2	-0.531074972353244	-0.531074972353244\\
0.256	0.2	-0.484488783085505	-0.484488783085505\\
0.272	0.2	-0.44129548609371	-0.44129548609371\\
0.288	0.2	-0.401380981167493	-0.401380981167493\\
0.304	0.2	-0.364604367608797	-0.364604367608797\\
0.32	0.2	-0.330806888051372	-0.330806888051372\\
0.336	0.2	-0.299818998561052	-0.299818998561052\\
0.352	0.2	-0.271465885745105	-0.271465885745105\\
0.368	0.2	-0.24557171239171	-0.24557171239171\\
0.384	0.2	-0.221962830856632	-0.221962830856632\\
0.4	0.2	-0.200470164097505	-0.200470164097505\\
0	0.24	-1.51840659961129	-1.51840659961129\\
0.016	0.24	-1.56618645121326	-1.56618645121326\\
0.032	0.24	-1.44284091350933	-1.44284091350933\\
0.048	0.24	-1.34036402549206	-1.34036402549206\\
0.064	0.24	-1.25300555227409	-1.25300555227409\\
0.08	0.24	-1.17170159164893	-1.17170159164893\\
0.096	0.24	-1.09312888981507	-1.09312888981507\\
0.112	0.24	-1.01659513276161	-1.01659513276161\\
0.128	0.24	-0.942325921452994	-0.942325921452994\\
0.144	0.24	-0.870776080409976	-0.870776080409976\\
0.16	0.24	-0.802376985841136	-0.802376985841136\\
0.176	0.24	-0.737457021125567	-0.737457021125567\\
0.192	0.24	-0.67622878602994	-0.67622878602994\\
0.208	0.24	-0.618800329455693	-0.618800329455693\\
0.224	0.24	-0.565193333466597	-0.565193333466597\\
0.24	0.24	-0.515361633064305	-0.515361633064305\\
0.256	0.24	-0.469207733916037	-0.469207733916037\\
0.272	0.24	-0.426596718423898	-0.426596718423898\\
0.288	0.24	-0.38736759946335	-0.38736759946335\\
0.304	0.24	-0.351342411712229	-0.351342411712229\\
0.32	0.24	-0.318333383346763	-0.318333383346763\\
0.336	0.24	-0.28814851538104	-0.28814851538104\\
0.352	0.24	-0.260595857548818	-0.260595857548818\\
0.368	0.24	-0.235486726795717	-0.235486726795717\\
0.384	0.24	-0.212638074184518	-0.212638074184518\\
0.4	0.24	-0.191874170588805	-0.191874170588805\\
0	0.28	-1.35531026114111	-1.35531026114111\\
0.016	0.28	-1.47448335380631	-1.47448335380631\\
0.032	0.28	-1.40978474730413	-1.40978474730413\\
0.048	0.28	-1.32693272920952	-1.32693272920952\\
0.064	0.28	-1.24267557366844	-1.24267557366844\\
0.08	0.28	-1.15849774139152	-1.15849774139152\\
0.096	0.28	-1.07560151925625	-1.07560151925625\\
0.112	0.28	-0.995067550805401	-0.995067550805401\\
0.128	0.28	-0.917698908439204	-0.917698908439204\\
0.144	0.28	-0.844036669111217	-0.844036669111217\\
0.16	0.28	-0.77441465899391	-0.77441465899391\\
0.176	0.28	-0.709009531236808	-0.709009531236808\\
0.192	0.28	-0.647879913773896	-0.647879913773896\\
0.208	0.28	-0.590996384682068	-0.590996384682068\\
0.224	0.28	-0.538264621054011	-0.538264621054011\\
0.24	0.28	-0.489543458001039	-0.489543458001039\\
0.256	0.28	-0.44465905081076	-0.44465905081076\\
0.272	0.28	-0.403415981754273	-0.403415981754273\\
0.288	0.28	-0.36560593432729	-0.36560593432729\\
0.304	0.28	-0.331014415373079	-0.331014415373079\\
0.32	0.28	-0.299425906225125	-0.299425906225125\\
0.336	0.28	-0.270627750174509	-0.270627750174509\\
0.352	0.28	-0.244413026114687	-0.244413026114687\\
0.368	0.28	-0.220582612231458	-0.220582612231458\\
0.384	0.28	-0.198946606213644	-0.198946606213644\\
0.4	0.28	-0.179325237806735	-0.179325237806735\\
0	0.32	-1.20155629761702	-1.20155629761702\\
0.016	0.32	-1.36442406213687	-1.36442406213687\\
0.032	0.32	-1.35378327622625	-1.35378327622625\\
0.048	0.32	-1.29110085147462	-1.29110085147462\\
0.064	0.32	-1.21078795131898	-1.21078795131898\\
0.08	0.32	-1.12469418490023	-1.12469418490023\\
0.096	0.32	-1.03853759477539	-1.03853759477539\\
0.112	0.32	-0.955135925875334	-0.955135925875334\\
0.128	0.32	-0.875832902446616	-0.875832902446616\\
0.144	0.32	-0.801226684379626	-0.801226684379626\\
0.16	0.32	-0.731532057736231	-0.731532057736231\\
0.176	0.32	-0.666758555248047	-0.666758555248047\\
0.192	0.32	-0.60680008998477	-0.60680008998477\\
0.208	0.32	-0.55148256747235	-0.55148256747235\\
0.224	0.32	-0.500591171999522	-0.500591171999522\\
0.24	0.32	-0.453887333775392	-0.453887333775392\\
0.256	0.32	-0.411120044161103	-0.411120044161103\\
0.272	0.32	-0.372033775411435	-0.372033775411435\\
0.288	0.32	-0.336374167338745	-0.336374167338745\\
0.304	0.32	-0.30389213515672	-0.30389213515672\\
0.32	0.32	-0.274346805555815	-0.274346805555815\\
0.336	0.32	-0.247507558664644	-0.247507558664644\\
0.352	0.32	-0.223155378971819	-0.223155378971819\\
0.368	0.32	-0.201083670605756	-0.201083670605756\\
0.384	0.32	-0.181098658997589	-0.181098658997589\\
0.4	0.32	-0.163019475996785	-0.163019475996785\\
0	0.36	-1.05796386772909	-1.05796386772909\\
0.016	0.36	-1.24426657422386	-1.24426657422386\\
0.032	0.36	-1.27851245268081	-1.27851245268081\\
0.048	0.36	-1.23393311695668	-1.23393311695668\\
0.064	0.36	-1.15741900559328	-1.15741900559328\\
0.08	0.36	-1.07011502864402	-1.07011502864402\\
0.096	0.36	-0.981796161555991	-0.981796161555991\\
0.112	0.36	-0.896785243303076	-0.896785243303076\\
0.128	0.36	-0.81685015368829	-0.81685015368829\\
0.144	0.36	-0.742587851151629	-0.742587851151629\\
0.16	0.36	-0.674064585253199	-0.674064585253199\\
0.176	0.36	-0.611107699988238	-0.611107699988238\\
0.192	0.36	-0.5534388791413	-0.5534388791413\\
0.208	0.36	-0.500736060858837	-0.500736060858837\\
0.224	0.36	-0.452663171773024	-0.452663171773024\\
0.24	0.36	-0.408885083123212	-0.408885083123212\\
0.256	0.36	-0.369075533219779	-0.369075533219779\\
0.272	0.36	-0.332921506710004	-0.332921506710004\\
0.288	0.36	-0.300125688383935	-0.300125688383935\\
0.304	0.36	-0.270407778455424	-0.270407778455424\\
0.32	0.36	-0.243505080696908	-0.243505080696908\\
0.336	0.36	-0.219172598966916	-0.219172598966916\\
0.352	0.36	-0.197182790456254	-0.197182790456254\\
0.368	0.36	-0.177325077010031	-0.177325077010031\\
0.384	0.36	-0.159405187948432	-0.159405187948432\\
0.4	0.36	-0.143244389480621	-0.143244389480621\\
0	0.4	-0.925116780127006	-0.925116780127006\\
0.016	0.4	-1.12063326041443	-1.12063326041443\\
0.032	0.4	-1.1875518943916	-1.1875518943916\\
0.048	0.4	-1.15671201145733	-1.15671201145733\\
0.064	0.4	-1.08297990571578	-1.08297990571578\\
0.08	0.4	-0.994965260566593	-0.994965260566593\\
0.096	0.4	-0.905627635665476	-0.905627635665476\\
0.112	0.4	-0.820384527360852	-0.820384527360852\\
0.128	0.4	-0.741240955681196	-0.741240955681196\\
0.144	0.4	-0.668709901496598	-0.668709901496598\\
0.16	0.4	-0.602673923462489	-0.602673923462489\\
0.176	0.4	-0.542764665865374	-0.542764665865374\\
0.192	0.4	-0.488527981998239	-0.488527981998239\\
0.208	0.4	-0.439494854146628	-0.439494854146628\\
0.224	0.4	-0.395211150458094	-0.395211150458094\\
0.24	0.4	-0.355249425528104	-0.355249425528104\\
0.256	0.4	-0.319212883045414	-0.319212883045414\\
0.272	0.4	-0.286735914830361	-0.286735914830361\\
0.288	0.4	-0.257483151554698	-0.257483151554698\\
0.304	0.4	-0.231147882419456	-0.231147882419456\\
0.32	0.4	-0.207450230296754	-0.207450230296754\\
0.336	0.4	-0.186135261283301	-0.186135261283301\\
0.352	0.4	-0.166971114462906	-0.166971114462906\\
0.368	0.4	-0.149747194587646	-0.149747194587646\\
0.384	0.4	-0.13427244949593	-0.13427244949593\\
0.4	0.4	-0.120373743271534	-0.120373743271534\\
0	0.44	-0.803368224752894	-0.803368224752894\\
0.016	0.44	-0.998373657583567	-0.998373657583567\\
0.032	0.44	-1.08402400629299	-1.08402400629299\\
0.048	0.44	-1.06078495877546	-1.06078495877546\\
0.064	0.44	-0.988172591078087	-0.988172591078087\\
0.08	0.44	-0.899828493009093	-0.899828493009093\\
0.096	0.44	-0.810683726195942	-0.810683726195942\\
0.112	0.44	-0.726697252204915	-0.726697252204915\\
0.128	0.44	-0.649870562858243	-0.649870562858243\\
0.144	0.44	-0.580533158061719	-0.580533158061719\\
0.16	0.44	-0.518346628213325	-0.518346628213325\\
0.176	0.44	-0.462736571435573	-0.462736571435573\\
0.192	0.44	-0.413074052382942	-0.413074052382942\\
0.208	0.44	-0.368748740222903	-0.368748740222903\\
0.224	0.44	-0.329195739958341	-0.329195739958341\\
0.24	0.44	-0.29390296430996	-0.29390296430996\\
0.256	0.44	-0.262410593636886	-0.262410593636886\\
0.272	0.44	-0.23430755104629	-0.23430755104629\\
0.288	0.44	-0.209227066731544	-0.209227066731544\\
0.304	0.44	-0.186842181113172	-0.186842181113172\\
0.32	0.44	-0.166861514067958	-0.166861514067958\\
0.336	0.44	-0.149025407251693	-0.149025407251693\\
0.352	0.44	-0.133102455744891	-0.133102455744891\\
0.368	0.44	-0.118886409726713	-0.118886409726713\\
0.384	0.44	-0.10619341470256	-0.10619341470256\\
0.4	0.44	-0.094859556296174	-0.094859556296174\\
0	0.48	-0.692854142573543	-0.692854142573543\\
0.016	0.48	-0.880629371739028	-0.880629371739028\\
0.032	0.48	-0.970309529665882	-0.970309529665882\\
0.048	0.48	-0.9474383401405	-0.9474383401405\\
0.064	0.48	-0.873950952379945	-0.873950952379945\\
0.08	0.48	-0.785659982396307	-0.785659982396307\\
0.096	0.48	-0.698018366865086	-0.698018366865086\\
0.112	0.48	-0.616881229128294	-0.616881229128294\\
0.128	0.48	-0.543975311747557	-0.543975311747557\\
0.144	0.48	-0.479340612813244	-0.479340612813244\\
0.16	0.48	-0.422382754895783	-0.422382754895783\\
0.176	0.48	-0.372315939280738	-0.372315939280738\\
0.192	0.48	-0.328342833315792	-0.328342833315792\\
0.208	0.48	-0.289722284599675	-0.289722284599675\\
0.224	0.48	-0.255790033683367	-0.255790033683367\\
0.24	0.48	-0.225960377234192	-0.225960377234192\\
0.256	0.48	-0.199720659964032	-0.199720659964032\\
0.272	0.48	-0.176623561832746	-0.176623561832746\\
0.288	0.48	-0.156279190231649	-0.156279190231649\\
0.304	0.48	-0.138347732545959	-0.138347732545959\\
0.32	0.48	-0.122532900840708	-0.122532900840708\\
0.336	0.48	-0.108576188905933	-0.108576188905933\\
0.352	0.48	-0.0962518826505247	-0.0962518826505247\\
0.368	0.48	-0.085362740728398	-0.085362740728398\\
0.384	0.48	-0.0757362603959763	-0.0757362603959763\\
0.4	0.48	-0.0672214500006735	-0.0672214500006735\\
0	0.52	-0.593513955561926	-0.593513955561926\\
0.016	0.52	-0.768967767060241	-0.768967767060241\\
0.032	0.52	-0.847858342801672	-0.847858342801672\\
0.048	0.52	-0.817813032076343	-0.817813032076343\\
0.064	0.52	-0.741493019921002	-0.741493019921002\\
0.08	0.52	-0.653776614881578	-0.653776614881578\\
0.096	0.52	-0.569079850447827	-0.569079850447827\\
0.112	0.52	-0.492477739153904	-0.492477739153904\\
0.128	0.52	-0.425147701073672	-0.425147701073672\\
0.144	0.52	-0.366739379069759	-0.366739379069759\\
0.16	0.52	-0.316374541859663	-0.316374541859663\\
0.176	0.52	-0.273057506117759	-0.273057506117759\\
0.192	0.52	-0.235834884692199	-0.235834884692199\\
0.208	0.52	-0.203850120404884	-0.203850120404884\\
0.224	0.52	-0.176354960482278	-0.176354960482278\\
0.24	0.52	-0.1527042595582	-0.1527042595582\\
0.256	0.52	-0.132345172530827	-0.132345172530827\\
0.272	0.52	-0.114805252308466	-0.114805252308466\\
0.288	0.52	-0.0996811898771022	-0.0996811898771022\\
0.304	0.52	-0.0866287713745487	-0.0866287713745487\\
0.32	0.52	-0.0753541536691316	-0.0753541536691316\\
0.336	0.52	-0.0656063791346627	-0.0656063791346627\\
0.352	0.52	-0.0571709917113671	-0.0571709917113671\\
0.368	0.52	-0.0498646073025763	-0.0498646073025763\\
0.384	0.52	-0.0435303013040048	-0.0435303013040048\\
0.4	0.52	-0.0380336917558641	-0.0380336917558641\\
0	0.56	-0.505117195484334	-0.505117195484334\\
0.016	0.56	-0.663479478168192	-0.663479478168192\\
0.032	0.56	-0.717107062783641	-0.717107062783641\\
0.048	0.56	-0.672871529059509	-0.672871529059509\\
0.064	0.56	-0.59218792258052	-0.59218792258052\\
0.08	0.56	-0.505844933369381	-0.505844933369381\\
0.096	0.56	-0.425694370949378	-0.425694370949378\\
0.112	0.56	-0.355389953635373	-0.355389953635373\\
0.128	0.56	-0.29531055583569	-0.29531055583569\\
0.144	0.56	-0.244631787306612	-0.244631787306612\\
0.16	0.56	-0.202175563002223	-0.202175563002223\\
0.176	0.56	-0.166746382835989	-0.166746382835989\\
0.192	0.56	-0.137253511890628	-0.137253511890628\\
0.208	0.56	-0.112745234139951	-0.112745234139951\\
0.224	0.56	-0.0924083693458891	-0.0924083693458891\\
0.24	0.56	-0.0755553286593104	-0.0755553286593104\\
0.256	0.56	-0.0616078653835072	-0.0616078653835072\\
0.272	0.56	-0.0500811234346288	-0.0500811234346288\\
0.288	0.56	-0.0405692561192618	-0.0405692561192618\\
0.304	0.56	-0.0327329312337652	-0.0327329312337652\\
0.32	0.56	-0.0262886642933627	-0.0262886642933627\\
0.336	0.56	-0.0209997918267031	-0.0209997918267031\\
0.352	0.56	-0.0166688668764332	-0.0166688668764332\\
0.368	0.56	-0.013131267928092	-0.013131267928092\\
0.384	0.56	-0.0102498347988744	-0.0102498347988744\\
0.4	0.56	-0.00791037008058515	-0.00791037008058515\\
0	0.6	-0.427294490970615	-0.427294490970615\\
0.016	0.6	-0.562783035680713	-0.562783035680713\\
0.032	0.6	-0.577509293921009	-0.577509293921009\\
0.048	0.6	-0.513421095229235	-0.513421095229235\\
0.064	0.6	-0.427638908176406	-0.427638908176406\\
0.08	0.6	-0.34386749913887	-0.34386749913887\\
0.096	0.6	-0.270041146450417	-0.270041146450417\\
0.112	0.6	-0.207850956216287	-0.207850956216287\\
0.128	0.6	-0.156680775329784	-0.156680775329784\\
0.144	0.6	-0.115176789690941	-0.115176789690941\\
0.16	0.6	-0.0818611405090089	-0.0818611405090089\\
0.176	0.6	-0.0553584422695373	-0.0553584422695373\\
0.192	0.6	-0.0344658258557819	-0.0344658258557819\\
0.208	0.6	-0.0181611976839709	-0.0181611976839709\\
0.224	0.6	-0.00558878341159714	-0.00558878341159714\\
0.24	0.6	0.00396206490101305	0.00396206490101305\\
0.256	0.6	0.0110784697113029	0.0110784697113029\\
0.272	0.6	0.0162437333995997	0.0162437333995997\\
0.288	0.6	0.0198544984121326	0.0198544984121326\\
0.304	0.6	0.0222353661486586	0.0222353661486586\\
0.32	0.6	0.0236512133127043	0.0236512133127043\\
0.336	0.6	0.0243175068591808	0.0243175068591808\\
0.352	0.6	0.0244089133002505	0.0244089133002505\\
0.368	0.6	0.0240664688140109	0.0240664688140109\\
0.384	0.6	0.023403541344006	0.023403541344006\\
0.4	0.6	0.0225107817200494	0.0225107817200494\\
0	0.64	-0.359571392139174	-0.359571392139174\\
0.016	0.64	-0.463937157646822	-0.463937157646822\\
0.032	0.64	-0.427679472571375	-0.427679472571375\\
0.048	0.64	-0.340191230663013	-0.340191230663013\\
0.064	0.64	-0.249681033055859	-0.249681033055859\\
0.08	0.64	-0.170167046790712	-0.170167046790712\\
0.096	0.64	-0.104619256870498	-0.104619256870498\\
0.112	0.64	-0.0523819494699283	-0.0523819494699283\\
0.128	0.64	-0.011723533499696	-0.011723533499696\\
0.144	0.64	0.0192572817942647	0.0192572817942647\\
0.16	0.64	0.0423188335533712	0.0423188335533712\\
0.176	0.64	0.0589858630330692	0.0589858630330692\\
0.192	0.64	0.0705418453637501	0.0705418453637501\\
0.208	0.64	0.0780504417201757	0.0780504417201757\\
0.224	0.64	0.0823849920481512	0.0823849920481512\\
0.24	0.64	0.0842579756887994	0.0842579756887994\\
0.256	0.64	0.0842476085382997	0.0842476085382997\\
0.272	0.64	0.0828208459504896	0.0828208459504896\\
0.288	0.64	0.0803528720453704	0.0803528720453704\\
0.304	0.64	0.0771434374518032	0.0771434374518032\\
0.32	0.64	0.0734304742550208	0.0734304742550208\\
0.336	0.64	0.0694014007527323	0.0694014007527323\\
0.352	0.64	0.0652024842198071	0.0652024842198071\\
0.368	0.64	0.0609465794426427	0.0609465794426427\\
0.384	0.64	0.0567195128459777	0.0567195128459777\\
0.4	0.64	0.052585339440015	0.052585339440015\\
0	0.68	-0.301403619267417	-0.301403619267417\\
0.016	0.68	-0.362320115489773	-0.362320115489773\\
0.032	0.68	-0.26564588596872	-0.26564588596872\\
0.048	0.68	-0.153957560412029	-0.153957560412029\\
0.064	0.68	-0.0604095449074061	-0.0604095449074061\\
0.08	0.68	0.0126328699473576	0.0126328699473576\\
0.096	0.68	0.0677936755952207	0.0677936755952207\\
0.112	0.68	0.108258498903978	0.108258498903978\\
0.128	0.68	0.136901849955536	0.136901849955536\\
0.144	0.68	0.156147209893126	0.156147209893126\\
0.16	0.68	0.167993199305529	0.167993199305529\\
0.176	0.68	0.174072770034615	0.174072770034615\\
0.192	0.68	0.175712730555133	0.175712730555133\\
0.208	0.68	0.173986192942496	0.173986192942496\\
0.224	0.68	0.169757044239258	0.169757044239258\\
0.24	0.68	0.163717154142693	0.163717154142693\\
0.256	0.68	0.156417334277452	0.156417334277452\\
0.272	0.68	0.148293024899499	0.148293024899499\\
0.288	0.68	0.139685565693094	0.139685565693094\\
0.304	0.68	0.130859780762386	0.130859780762386\\
0.32	0.68	0.122018493080322	0.122018493080322\\
0.336	0.68	0.113314484364363	0.113314484364363\\
0.352	0.68	0.104860332038072	0.104860332038072\\
0.368	0.68	0.0967364838965129	0.0967364838965129\\
0.384	0.68	0.0889978714480643	0.0889978714480643\\
0.4	0.68	0.0816793129214929	0.0816793129214929\\
0	0.72	-0.252212500030056	-0.252212500030056\\
0.016	0.72	-0.251588121153769	-0.251588121153769\\
0.032	0.72	-0.0892013565955704	-0.0892013565955704\\
0.048	0.72	0.0443013981199134	0.0443013981199134\\
0.064	0.72	0.137787195379647	0.137787195379647\\
0.08	0.72	0.201632612092003	0.201632612092003\\
0.096	0.72	0.244190774820231	0.244190774820231\\
0.112	0.72	0.271133041640457	0.271133041640457\\
0.128	0.72	0.286402829585654	0.286402829585654\\
0.144	0.72	0.292874943225351	0.292874943225351\\
0.16	0.72	0.292727222958779	0.292727222958779\\
0.176	0.72	0.287649855567924	0.287649855567924\\
0.192	0.72	0.278970927414804	0.278970927414804\\
0.208	0.72	0.267738502749173	0.267738502749173\\
0.224	0.72	0.254778824574949	0.254778824574949\\
0.24	0.72	0.240740194536124	0.240740194536124\\
0.256	0.72	0.226127388443276	0.226127388443276\\
0.272	0.72	0.211329269459802	0.211329269459802\\
0.288	0.72	0.196641209581941	0.196641209581941\\
0.304	0.72	0.182283397564025	0.182283397564025\\
0.32	0.72	0.168415817737501	0.168415817737501\\
0.336	0.72	0.155150504376813	0.155150504376813\\
0.352	0.72	0.142561554371008	0.142561554371008\\
0.368	0.72	0.130693291309711	0.130693291309711\\
0.384	0.72	0.119566904428178	0.119566904428178\\
0.4	0.72	0.109185829905326	0.109185829905326\\
0	0.76	-0.211419589368306	-0.211419589368306\\
0.016	0.76	-0.123858471152437	-0.123858471152437\\
0.032	0.76	0.103669328495047	0.103669328495047\\
0.048	0.76	0.253232714975868	0.253232714975868\\
0.064	0.76	0.342197907817239	0.342197907817239\\
0.08	0.76	0.393688715364012	0.393688715364012\\
0.096	0.76	0.421392825432802	0.421392825432802\\
0.112	0.76	0.433190566846874	0.433190566846874\\
0.128	0.76	0.433919529697279	0.433919529697279\\
0.144	0.76	0.426791266149145	0.426791266149145\\
0.16	0.76	0.414082753971023	0.414082753971023\\
0.176	0.76	0.397482208429846	0.397482208429846\\
0.192	0.76	0.378273489528349	0.378273489528349\\
0.208	0.76	0.357443832213801	0.357443832213801\\
0.224	0.76	0.335753372988847	0.335753372988847\\
0.24	0.76	0.313784106249959	0.313784106249959\\
0.256	0.76	0.291976606797547	0.291976606797547\\
0.272	0.76	0.270658692269575	0.270658692269575\\
0.288	0.76	0.250068301113826	0.250068301113826\\
0.304	0.76	0.230371962834483	0.230371962834483\\
0.32	0.76	0.211679783643267	0.211679783643267\\
0.336	0.76	0.194057620052297	0.194057620052297\\
0.352	0.76	0.177536958914967	0.177536958914967\\
0.368	0.76	0.162122917564261	0.162122917564261\\
0.384	0.76	0.14780070036428	0.14780070036428\\
0.4	0.76	0.13454078782382	0.13454078782382\\
0	0.8	-0.178479727036902	-0.178479727036902\\
0.016	0.8	0.0297387842736794	0.0297387842736794\\
0.032	0.8	0.314317468090502	0.314317468090502\\
0.048	0.8	0.470943894448076	0.470943894448076\\
0.064	0.8	0.5497699764623	0.5497699764623\\
0.08	0.8	0.585455741221746	0.585455741221746\\
0.096	0.8	0.596112212613852	0.596112212613852\\
0.112	0.8	0.591335973036893	0.591335973036893\\
0.128	0.8	0.576593576244887	0.576593576244887\\
0.144	0.8	0.555280979887861	0.555280979887861\\
0.16	0.8	0.529678797836972	0.529678797836972\\
0.176	0.8	0.501408187709876	0.501408187709876\\
0.192	0.8	0.471661483724177	0.471661483724177\\
0.208	0.8	0.441329258157913	0.441329258157913\\
0.224	0.8	0.411077757430799	0.411077757430799\\
0.24	0.8	0.381400897383578	0.381400897383578\\
0.256	0.8	0.352657948587282	0.352657948587282\\
0.272	0.8	0.325102283619813	0.325102283619813\\
0.288	0.8	0.298903977866729	0.298903977866729\\
0.304	0.8	0.274167863599542	0.274167863599542\\
0.32	0.8	0.250948057167061	0.250948057167061\\
0.336	0.8	0.229259673782831	0.229259673782831\\
0.352	0.8	0.209088266434119	0.209088266434119\\
0.368	0.8	0.190397409998269	0.190397409998269\\
0.384	0.8	0.173134769550317	0.173134769550317\\
0.4	0.8	0.15723692947351	0.15723692947351\\
0	0.84	-0.152912061123267	-0.152912061123267\\
0.016	0.84	0.218050249608346	0.218050249608346\\
0.032	0.84	0.54295778768465	0.54295778768465\\
0.048	0.84	0.694866182797795	0.694866182797795\\
0.064	0.84	0.757110315709758	0.757110315709758\\
0.08	0.84	0.773439323077205	0.773439323077205\\
0.096	0.84	0.76502292691584	0.76502292691584\\
0.112	0.84	0.742502786144416	0.742502786144416\\
0.128	0.84	0.711637667517783	0.711637667517783\\
0.144	0.84	0.675827399139429	0.675827399139429\\
0.16	0.84	0.637250422086164	0.637250422086164\\
0.176	0.84	0.597392837447219	0.597392837447219\\
0.192	0.84	0.55730824893285	0.55730824893285\\
0.208	0.84	0.517755987595724	0.517755987595724\\
0.224	0.84	0.479282264176951	0.479282264176951\\
0.24	0.84	0.442272843139645	0.442272843139645\\
0.256	0.84	0.40699021725453	0.40699021725453\\
0.272	0.84	0.373601428145898	0.373601428145898\\
0.288	0.84	0.342199643726225	0.342199643726225\\
0.304	0.84	0.312821219369736	0.312821219369736\\
0.32	0.84	0.285459308678419	0.285459308678419\\
0.336	0.84	0.260074750618014	0.260074750618014\\
0.352	0.84	0.236604768275425	0.236604768275425\\
0.368	0.84	0.214969894014577	0.214969894014577\\
0.384	0.84	0.195079452236075	0.195079452236075\\
0.4	0.84	0.17683586854982	0.17683586854982\\
0	0.88	-0.134328835171164	-0.134328835171164\\
0.016	0.88	0.448447964617592	0.448447964617592\\
0.032	0.88	0.78824819323317	0.78824819323317\\
0.048	0.88	0.921667096886595	0.921667096886595\\
0.064	0.88	0.96051203142889	0.96051203142889\\
0.08	0.88	0.954061196515313	0.954061196515313\\
0.096	0.88	0.924835237115257	0.924835237115257\\
0.112	0.88	0.883726485013241	0.883726485013241\\
0.128	0.88	0.836403602379909	0.836403602379909\\
0.144	0.88	0.786073966803991	0.786073966803991\\
0.16	0.88	0.734703964853549	0.734703964853549\\
0.176	0.88	0.683577115049943	0.683577115049943\\
0.192	0.88	0.633563198655981	0.633563198655981\\
0.208	0.88	0.585258299323386	0.585258299323386\\
0.224	0.88	0.539065009744168	0.539065009744168\\
0.24	0.88	0.495243250929919	0.495243250929919\\
0.256	0.88	0.453945411143107	0.453945411143107\\
0.272	0.88	0.41524222575296	0.41524222575296\\
0.288	0.88	0.379142601126893	0.379142601126893\\
0.304	0.88	0.345609127848716	0.345609127848716\\
0.32	0.88	0.314570339292205	0.314570339292205\\
0.336	0.88	0.285930423017534	0.285930423017534\\
0.352	0.88	0.259576899058841	0.259576899058841\\
0.368	0.88	0.235386659752701	0.235386659752701\\
0.384	0.88	0.213230684221673	0.213230684221673\\
0.4	0.88	0.192977680454041	0.192977680454041\\
0	0.92	-0.122461989497086	-0.122461989497086\\
0.016	0.92	0.725102498831703	0.725102498831703\\
0.032	0.92	1.04697915890235	1.04697915890235\\
0.048	0.92	1.14722921587089	1.14722921587089\\
0.064	0.92	1.15601110997042	1.15601110997042\\
0.08	0.92	1.12373590355027	1.12373590355027\\
0.096	0.92	1.07237315682125	1.07237315682125\\
0.112	0.92	1.01221623833569	1.01221623833569\\
0.128	0.92	0.948446483775488	0.948446483775488\\
0.144	0.92	0.88388087639572	0.88388087639572\\
0.16	0.92	0.820166651940244	0.820166651940244\\
0.176	0.92	0.758321255907947	0.758321255907947\\
0.192	0.92	0.698989689908949	0.698989689908949\\
0.208	0.92	0.64257661513027	0.64257661513027\\
0.224	0.92	0.589320835529854	0.589320835529854\\
0.24	0.92	0.539341722544528	0.539341722544528\\
0.256	0.92	0.492670826394566	0.492670826394566\\
0.272	0.92	0.44927484552039	0.44927484552039\\
0.288	0.92	0.409073011326445	0.409073011326445\\
0.304	0.92	0.371950539352151	0.371950539352151\\
0.32	0.92	0.337769136109881	0.337769136109881\\
0.336	0.92	0.306375218139879	0.306375218139879\\
0.352	0.92	0.277606316659642	0.277606316659642\\
0.368	0.92	0.251296029025816	0.251296029025816\\
0.384	0.92	0.227277802229029	0.227277802229029\\
0.4	0.92	0.205387777894253	0.205387777894253\\
0	0.96	-0.117187856305603	-0.117187856305603\\
0.016	0.96	1.04728934325923	1.04728934325923\\
0.032	0.96	1.31392631264612	1.31392631264612\\
0.048	0.96	1.36670636602733	1.36670636602733\\
0.064	0.96	1.33947440722287	1.33947440722287\\
0.08	0.96	1.27895758865462	1.27895758865462\\
0.096	0.96	1.20465236244949	1.20465236244949\\
0.112	0.96	1.1254226640589	1.1254226640589\\
0.128	0.96	1.04558290191132	1.04558290191132\\
0.144	0.96	0.967374764993555	0.967374764993555\\
0.16	0.96	0.89202894366797	0.89202894366797\\
0.176	0.96	0.820240831353184	0.820240831353184\\
0.192	0.96	0.752395723058551	0.752395723058551\\
0.208	0.96	0.688683644120513	0.688683644120513\\
0.224	0.96	0.62916358214215	0.62916358214215\\
0.24	0.96	0.573803141412979	0.573803141412979\\
0.256	0.96	0.522505251812147	0.522505251812147\\
0.272	0.96	0.475127347601998	0.475127347601998\\
0.288	0.96	0.431495698898808	0.431495698898808\\
0.304	0.96	0.391416345327067	0.391416345327067\\
0.32	0.96	0.354683500377829	0.354683500377829\\
0.336	0.96	0.321086002471492	0.321086002471492\\
0.352	0.96	0.290412227317075	0.290412227317075\\
0.368	0.96	0.262453777131812	0.262453777131812\\
0.384	0.96	0.237008195110552	0.237008195110552\\
0.4	0.96	0.213880904258451	0.213880904258451\\
0	1	-0.118550430695653	-0.118550430695653\\
0.016	1	1.40812543576314	1.40812543576314\\
0.032	1	1.58190701980675	1.58190701980675\\
0.048	1	1.57466070106659	1.57466070106659\\
0.064	1	1.50671686765237	1.50671686765237\\
0.08	1	1.41639399516164	1.41639399516164\\
0.096	1	1.31895581664366	1.31895581664366\\
0.112	1	1.22109922000335	1.22109922000335\\
0.128	1	1.12594107708658	1.12594107708658\\
0.144	1	1.03498979256607	1.03498979256607\\
0.16	1	0.948978216820908	0.948978216820908\\
0.176	1	0.868234349775666	0.868234349775666\\
0.192	1	0.792856526972144	0.792856526972144\\
0.208	1	0.72280282491717	0.72280282491717\\
0.224	1	0.65794110959459	0.65794110959459\\
0.24	1	0.598079868241119	0.598079868241119\\
0.256	1	0.542988835789605	0.542988835789605\\
0.272	1	0.492413633112364	0.492413633112364\\
0.288	1	0.446086526251275	0.446086526251275\\
0.304	1	0.403734460890808	0.403734460890808\\
0.32	1	0.365085072883738	0.365085072883738\\
0.336	1	0.329871144467088	0.329871144467088\\
0.352	1	0.297833846221884	0.297833846221884\\
0.368	1	0.268725024058469	0.268725024058469\\
0.384	1	0.242308735028794	0.242308735028794\\
0.4	1	0.218362194752444	0.218362194752444\\
};
\end{axis}

\begin{axis}[%
width=0.8\linewidth,
height=4cm,
at={(0,0)},
scale only axis,
xmin=0,
xmax=1,
ymin=0,
ymax=1,
axis line style={draw=none},
ticks=none,
axis x line*=bottom,
axis y line*=left
]
\end{axis}
\end{tikzpicture}%